\documentclass[10pt,twocolumn,reprint]{revtex4-1}
\usepackage[T1]{fontenc}
\usepackage[latin9]{inputenc}
\usepackage{amsmath}
\usepackage{amssymb}
\usepackage{graphicx}

\makeatletter

\providecommand{\tabularnewline}{\\}

\usepackage[caption=false]{subfig}

\makeatother

\begin{document}

\title{Modified belief propagation decoders for quantum low-density parity-check
codes}

\author{Alex Rigby}
\email{alex.rigby@utas.edu.au}

\author{JC Olivier}

\author{Peter Jarvis}

\affiliation{College of Sciences and Engineering, University of Tasmania, Hobart,
Tasmania 7005, Australia }
\begin{abstract}
Quantum low-density parity-check codes can be decoded using a syndrome
based $\mathrm{GF}(4)$ belief propagation decoder. However, the performance
of this decoder is limited both by unavoidable $4$-cycles in the
code's factor graph and the degenerate nature of quantum errors. For
the subclass of CSS codes, the number of $4$-cycles can be reduced
by breaking an error into an $X$ and $Z$ component and decoding
each with an individual $\mathrm{GF}(2)$ based decoder. However,
this comes at the expense of ignoring potential correlations between
these two error components. We present a number of modified belief
propagation decoders that address these issues. We propose a $\mathrm{GF}(2)$
based decoder for CSS codes that reintroduces error correlations by
reattempting decoding with adjusted error probabilities. We also propose
the use of an augmented decoder, which has previously been suggested
for classical binary low-density parity-check codes. This decoder
iteratively reattempts decoding on factor graphs that have a subset
of their check nodes duplicated. The augmented decoder can be based
on a $\mathrm{GF}(4)$ decoder for any code, a $\mathrm{GF}(2)$ decoder
for CSS code, or even a supernode decoder for a dual-containing CSS
code. For CSS codes, we further propose a $\mathrm{GF}(2)$ based
decoder that combines the augmented decoder with error probability
adjustment. We demonstrate the performance of these new decoders on
a range of different codes, showing that they perform favorably compared
to other decoders presented in literature.
\end{abstract}
\maketitle

\section{Introduction}

In the classical setting, low-density parity-check (LDPC) codes are
effective at protecting information against noise. LDPC codes are
particularly useful as their sparse structure permits the use of an
iterative belief propagation decoder that is of relatively low complexity
\citep{gallager1962low,mackay1999good}. Belief propagation is a message
passing algorithm that takes place on a code's factor graph. This
is a bipartite graph defined by a parity-check matrix for the code,
with each row corresponding to a check node and each column to an
error node. Quantum LDPC (QLDPC) codes, which are stabilizer codes
with sparse generators, can be used to protect against the effects
of a noisy quantum channel. The generators of an $n$-qubit stabilizer
code can be represented as elements of $\mathrm{GF}(4)^{n}$ \citep{calderbank1996quantum,gottesman1997stabilizer}.
This representation can be used to define a $\mathrm{GF}(4)$ parity-check
matrix, which allows for slightly altered $\mathrm{GF}(4)$ belief
propagation decoding of QLDPC codes \citep{babar2015fifteen}. The
requirement that all stabilizer generators must commute results in
unavoidable $4$-cycles in the factor graph associated with the $\mathrm{GF}(4)$
parity-check matrix \citep{poulin2008iterative}, which can be detrimental
to decoding performance \citep{mcgowan2003loop}. Belief propagation
performance is also limited by the fact that it attempts to converge
to the single most likely error (in a symbol-wise fashion), rather
than accounting for the degenerate nature of quantum errors \citep{poulin2008iterative}.
For the subclass of Calderbank-Shor-Steane (CSS) codes, the number
of $4$-cycles can be reduced by instead representing generators as
elements of $\mathrm{GF}(2)^{2n}$ \citep{calderbank1997quantum,gottesman1997stabilizer}.
This allows an error to be broken into an $X$ and $Z$ component,
which can then be decoded individually using two $\mathrm{GF}(2)$
belief propagation decoders \citep{mackay2004sparse}. However, for
many channels, including the depolarizing channel, this has the effect
of ignoring correlations between the two components \citep{babar2015fifteen}.

Modified belief propagation based decoders have been proposed that
aim to improve QLDPC decoding performance. Several decoders are presented
in Ref. \citep{poulin2008iterative} that aim to alleviate so-called
symmetric degeneracy errors, which occur as a result of symbol-wise
decoding in the face of error degeneracy. The best performing of these
is the random perturbation decoder, which attempts to break decoding
symmetries by iteratively reattempting decoding with randomly modified
channel error probabilities. The enhanced feedback (EFB) decoder of
Ref. \citep{wang2012enhanced} behaves similarly in that it also iteratively
reattempts decoding with modified error probabilities. However, unlike
the random perturbation decoder, this modification is informed by
the decoder's output. The supernode decoder of Ref. \citep{babar2015fifteen}
is a modification to the standard $\mathrm{GF}(4)$ decoder for the
subclass of dual-containing CSS codes. For this decoder, pairs of
check nodes in the factor graph are combined to form supernodes. This
both reduces decoding complexity and lowers the number of $4$-cycles
in the factor graph, which can lead to improved decoding performance.

The augmented decoder that we investigate has been previously proposed
for classical binary LDPC codes in Ref. \citep{rigby2018augmented}.
Like the random perturbation and EFB decoders, it also iteratively
reattempts decoding. Each of these attempts employs a version of the
standard factor graph with a randomly selected subset of check nodes
duplicated. In the classical case, this simple approach gives performance
that compares favorably with other, typically more complicated, decoders
presented in literature. In this paper we show that augmented decoders
can be applied to QLDPC codes whether the underlying decoder is $\mathrm{GF}(2)$,
$\mathrm{GF}(4)$, or supernode based. For CSS codes we propose the
$\mathrm{GF}(2)$ based adjusted decoder, which attempts to reintroduce
correlations between the $X$ and $Z$ components of an error that
are lost when using a standard $\mathrm{GF}(2)$ decoder. If one of
the two constituent $\mathrm{GF}(2)$ decoders fail, then the adjusted
decoder reattempts decoding of this component using error probabilities
that are modified according to the output of the other constituent
decoder (this is a slight generalization of the decoder presented
in Ref. \citep{delfosse2014decoding}). We also present a $\mathrm{GF(2)}$
based decoder for CSS codes that combines the augmented and adjusted
decoders. We simulate the performance of our decoders on six different
codes: two dual-containing CSS codes, two non-dual-containing CSS
codes, and two non-CSS codes. We show that for dual-containing CSS
codes our augmented $\mathrm{GF}(4)$, augmented supernode, and combined
decoders all outperform random perturbation and EFB decoders. For
the four other codes, we demonstrate that augmented $\mathrm{GF}(4)$
and supernode decoders perform similarly to to the random perturbation
and EFB decoders.

The paper is organized as follows. Sec. \ref{sec:Background} gives
an overview of belief propagation decoding for classical LDPC codes
and extends this to the quantum case. Sec. \ref{sec:Modified-decoders}
details the operation of existing modified decoders (random perturbation,
EFB, and supernode) and describes the adjusted, augmented, and combined
decoders that we propose. Sec. \ref{sec:Simulation-results} presents
simulation results for our decoders on six different codes, comparing
them to existing decoders. The paper is concluded in Sec. \ref{sec:Conclusion}.

\section{Background\label{sec:Background}}

\subsection{Classical codes}

A classical channel is the map $\Phi:\mathcal{A}_{x}\rightarrow\mathcal{A}_{y}$,
where $\mathcal{A}_{x}$ is the set of possible inputs and $\mathcal{A}_{y}$
is the set of possible outputs. We are concerned with channels where
the input and output sets are finite fields with $q$ elements; that
is, $\mathcal{A}_{x}=\mathcal{A}_{y}=\mathrm{GF}(q)$. In this case
the action of the channel can be expressed as
\begin{equation}
\Phi(x)=x+e=y,
\end{equation}
where $x\in\mathrm{GF}(q)$ is the channel input, $y\in\mathrm{GF}(q)$
is the channel output, and $e\in\mathrm{GF}(q)$ is an error (or noise)
symbol that occurs with probability $P(e)$. A channel $\Phi$ is
called symmetric if $P(0)=1-p$ and $P(e_{i})=p/(q-1)$ for $e_{i}\neq0$.
A code $\mathcal{C}\subseteq\mathrm{GF}(q)^{n}$ can be used to protect
against the noise introduced by the channel. Elements $\boldsymbol{x}\in\mathcal{C}$,
called codewords, are transmitted as $n$ sequential uses of $\Phi$
or, equivalently, as a single use of the combined channel $\Phi^{n}$,
which is comprised of $n$ copies of $\Phi$. The action of $\Phi^{n}$
on some input $\boldsymbol{x}\in\mathcal{C}$ is
\begin{equation}
\Phi^{n}(\boldsymbol{x})=\boldsymbol{x}+\boldsymbol{e}=\boldsymbol{y},
\end{equation}
where $\boldsymbol{y}\in\mathrm{GF}(q)^{n}$ is the channel output
and $\boldsymbol{e}\in\mathrm{GF}(q)^{n}$ is an error ``vector''.
Assuming the error components are independent, the probability of
an error $\boldsymbol{e}=(e_{1},\dots,e_{n})$ occurring is
\begin{equation}
P(\boldsymbol{e})=\prod_{i=1}^{n}P(e_{i}),\label{eq:classical error probability}
\end{equation}
where $P(e_{i})$ is the probability of the error symbol $e_{i}$
occurring on $\Phi$. The weight of a codeword $\boldsymbol{x}\in\mathcal{C}$
or an error $\boldsymbol{e}\in\mathrm{GF}(q)^{n}$ is the number of
non-zero components it contains. It follows from Eq. (\ref{eq:classical error probability})
that if $\Phi$ is symmetric, then the probability of $\boldsymbol{e}\in\mathrm{GF}(q)^{n}$
occurring depends only on its weight. The distance between two codewords
$\boldsymbol{x}_{i},\boldsymbol{x}_{j}\in\mathcal{C}$, denoted $\Delta(\boldsymbol{x}_{i},\boldsymbol{x}_{j})$,
is the number of components in which they differ. The distance of
$\mathcal{C}$ is 
\begin{equation}
d=\min_{\boldsymbol{x}_{i},\boldsymbol{x}_{j}\in\mathcal{C}}\Delta(\boldsymbol{x}_{i},\boldsymbol{x}_{j}).
\end{equation}
Equivalently, the distance of $\mathcal{C}$ is equal to the weight
of the lowest weight error that maps one codeword to another.

If a code $\mathcal{C}\subseteq\mathrm{GF}(q)^{n}$ forms an (additive)
group, then it is called additive; if it forms a vector space, then
it is called linear (note that there is no distinction between additive
and linear codes in the binary case). Suppose a linear code $\mathcal{C}$
has a basis $\mathcal{B}=\{\boldsymbol{b}_{1},\dots,\boldsymbol{b}_{k}\}$.
This defines a generator matrix
\begin{equation}
G^{T}=\left(\begin{array}{ccc}
\boldsymbol{b}_{1} & \cdots & \boldsymbol{b}_{k}\end{array}\right),
\end{equation}
where the basis elements are considered as column vectors. A generator
matrix can be defined in the same way for an additive code; however,
in this case $\mathcal{B}$ is a generating set. For a linear code,
the generator matrix defines a bijective encoding operation that maps
some $\boldsymbol{d}\in\mathrm{GF}(q)^{k}$ to a codeword $\boldsymbol{x}=G^{T}\boldsymbol{d}\in\mathcal{C}$
($\boldsymbol{d}$ is also considered as a column vector). A linear
code can also be defined as the kernel of a $\mathrm{GF}(q)$ parity-check
matrix $H$; that is,
\begin{equation}
\mathcal{C}=\{\boldsymbol{x}\in GF(q)^{n}:H\boldsymbol{x}=\boldsymbol{0}\}.
\end{equation}
Note that for a given code, neither the generator or parity-check
matrix is unique. If $H$ has $m$ rows, then $\dim(\mathcal{C})=k\geq n-m$,
with equality when $H$ is full rank. If $\mathcal{C}$ is linear
with dimension $k$ and distance $d$, then it is called an $[n,k]_{q}$
or $[n,k,d]_{q}$ code (the $q$ is typically omitted for binary codes,
where $q=2$). For a linear code, this distance is equal to weight
of the minimum weight non-zero codeword (as the errors that map one
codeword to another are the nontrivial $\boldsymbol{e}\in\mathcal{C}$).
The rate of a code is given by $R=k/n$.

The dual code of some code $\mathcal{C}\subseteq\mathrm{GF}(q)^{n}$
with respect to the inner product $\langle\cdot,\cdot\rangle:\mathrm{GF}(q)^{n}\times\mathrm{GF}(q)^{n}\rightarrow\mathrm{GF}(q)$
is
\begin{equation}
\mathcal{C}^{\perp}=\{\boldsymbol{c}\in\mathrm{GF}(q)^{n}:\langle\boldsymbol{c},\boldsymbol{x}\rangle=0\,\,\forall\,\,\boldsymbol{x}\in\mathcal{C}\}.
\end{equation}
$\mathcal{C}^{\perp}$ is the annihilator of $\mathcal{C}$ and is
therefore a linear code. If $\mathcal{C}^{\perp}\subseteq\mathcal{C}$,
then $\mathcal{C}$ is called dual-containing; if $\mathcal{C}\subseteq\mathcal{C}^{\perp}$,
then $\mathcal{C}$ is called self-orthogonal; and if $\mathcal{C}^{\perp}=\mathcal{C}$,
then $\mathcal{C}$ is called self-dual. Unless otherwise specified,
the dual code is with respect to the Euclidean inner product 
\begin{equation}
\langle\boldsymbol{c},\boldsymbol{x}\rangle=\boldsymbol{c}\cdot\boldsymbol{x}=\sum_{i=1}^{n}c_{i}x_{i}.
\end{equation}
In this case, if $\mathcal{C}$ is linear with generator matrix $G$,
then a necessary and sufficient condition for $\boldsymbol{c}\in\mathcal{C}^{\perp}$
is \textbf{$G\boldsymbol{c}=\boldsymbol{0}$}; that is, a generator
matrix for $\mathcal{C}$ is a parity-check matrix for $\mathcal{C}^{\perp}$.
Conversely, if $H$ is a parity-check matrix for $\mathcal{C}$, then
it is a generator matrix for $\mathcal{C}^{\perp}$.

The aim of a decoder is to determine the channel's input given its
output. For a linear code $\mathcal{C}$, this decoder can make use
of the error syndrome. If $\mathcal{C}$ has an $m\times n$ parity-check
matrix $H$ and the channel output is $\boldsymbol{y}$, then the
syndrome is
\begin{equation}
\boldsymbol{z}=H\boldsymbol{y}=H(\boldsymbol{x}+\boldsymbol{e})=H\boldsymbol{e}\in\mathrm{GF}(q)^{m}.\label{eq:classical syndrome definition}
\end{equation}
An optimal decoder returns the most probable error given the syndrome
measurement
\begin{equation}
\hat{\boldsymbol{e}}=\underset{\boldsymbol{e}\in\mathrm{GF}(q)^{n}}{\mathrm{argmax}}P(\boldsymbol{e}|\boldsymbol{z})=\underset{\boldsymbol{e}\in\mathrm{GF}(q)^{n}}{\mathrm{argmax}}P(\boldsymbol{e})\delta(H\boldsymbol{e}=\boldsymbol{z}),\label{eq:syndrome decoder}
\end{equation}
where $\delta(H\boldsymbol{e}=\boldsymbol{z})=1$ if $H\boldsymbol{e}=\boldsymbol{z}$
and $0$ otherwise. The channel input can then be estimated as $\hat{\boldsymbol{x}}=\boldsymbol{y}-\hat{\boldsymbol{e}}$.
If $\hat{\boldsymbol{e}}=\boldsymbol{e}$ (and hence $\hat{\boldsymbol{x}}=\boldsymbol{x}$),
then decoding is successful; otherwise, a decoding error has occurred.
Unfortunately, even in the simple case of a binary code operating
on the binary symmetric channel (a symmetric channel with $q=2$),
this decoding problem can be shown to be NP-complete \citep{berlekamp1978inherent}. 

It follows from Eq. (\ref{eq:classical syndrome definition}) that
the syndrome resulting from some error $\boldsymbol{e}\in\mathrm{GF}(q)^{n}$
depends only on which coset of $\mathrm{GF}(q)^{n}/\mathcal{C}$ it
belongs to. If $\hat{\boldsymbol{e}}$ is the most probable error
in the coset $\boldsymbol{e}+\mathcal{C}$, then the probability of
a decoding failure given the syndrome $\boldsymbol{z}=H\boldsymbol{e}$
is 
\begin{equation}
P(\boldsymbol{e}\neq\hat{\boldsymbol{e}}|\boldsymbol{z})=\frac{P(\boldsymbol{e}+\mathcal{C})-P(\hat{\boldsymbol{e}})}{P(\boldsymbol{e}+\mathcal{C})},
\end{equation}
where $P(\boldsymbol{e}+\mathcal{C})$ is the probability of any error
in $\boldsymbol{e}+\mathcal{C}$ occurring. Therefore, the probability
of a decoding error is high if the error probability distribution
over $\boldsymbol{e}+\mathcal{C}$ is not sharply peaked (that is,
if $P(\hat{\boldsymbol{e}})$ is small). If the channel is symmetric,
then this corresponds to $\boldsymbol{e}+\mathcal{C}$ containing
errors with similar weight to $\hat{\boldsymbol{e}}$, which will
be the case if $\mathcal{C}$ contains low weight codewords. It therefore
follows that the distance of $\mathcal{C}$ gives some indication
of the fraction of transmissions that will not be decoded correctly,
which is called call the frame error rate (FER).

\subsection{Factor graphs and belief propagation\label{subsec:LDPC-codes-and}}

\begin{figure}
\includegraphics[scale=0.55]{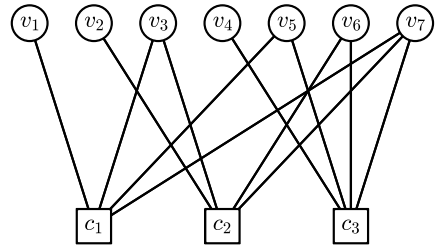}\caption{\label{hamming factor graph}The factor graph of the $[7,4,3]$ Hamming
code corresponding to the parity-check matrix given in Eq. (\ref{eq:hamming PCM}).
Error nodes are represented as circles and check nodes as squares.}
\end{figure}

The factor graph of a linear code is a bipartite graph $G=(V,C,E)$.
The error nodes $V=\{v_{1},\dots,v_{n}\}$ correspond to the $n$
error components, and the check nodes $C=\{c_{1},\dots,c_{m}\}$ correspond
to the $m$ constraints imposed by the rows of a parity-check matrix
$H$. An edge $\{c_{i},v_{j}\}\in E$ connects check node $c_{i}$
to error node $v_{j}$ if $H_{ij}\neq0$. For example, the $[7,4,3]$
Hamming code of Ref. \citep{hamming1950error} can be defined by the
parity-check matrix
\begin{equation}
H=\left(\begin{array}{ccccccc}
1 & 0 & 1 & 0 & 1 & 0 & 1\\
0 & 1 & 1 & 0 & 0 & 1 & 1\\
0 & 0 & 0 & 1 & 1 & 1 & 1
\end{array}\right),\label{eq:hamming PCM}
\end{equation}
which gives the factor graph shown in Fig. \ref{hamming factor graph}.
In general a given code does not have a unique factor graph as the
parity-check matrix from which it is defined is not unique. Furthermore,
except in the case of a binary code, the mapping from a parity-check
matrix to its corresponding factor graph is not one-to-one as an edge
only indicates that $H_{ij}\neq0$, it does not give the value of
$H_{ij}$ (although this information can be included by decorating
the edges). A walk is a sequence whose elements alternate between
connected nodes and the edges that connect them. The length of a walk
is the number of edges it contains. A path is a walk containing no
repeated nodes or edges with the exception that the first and last
node can be the same, in which case the path is called a cycle. The
bipartite nature of a code's factor graph ensures that the size of
all cycles is even and greater than or equal to four. As an example,
the walk $c_{1},\{c_{1},v_{5}\},v_{5},\{c_{3},v_{5}\},c_{3},\{c_{3},v_{7}\},v_{7},\{c_{1},v_{7}\},c_{1}$
in the graph of Fig. \ref{hamming factor graph} is a $4$-cycle (that
is, a cycle of length four). Typically a code's factor graph will
not be cycle free (that is, it will not be a tree) as if a code has
such a representation, then its distance is bounded by \citep{etzion1999codes}
\begin{equation}
d\leq\left\lfloor \frac{n}{k+1}\right\rfloor +\left\lfloor \frac{n+1}{k+1}\right\rfloor .
\end{equation}
For $R\geq1/2$ this reduces to $d\leq2$, and for $R>1/2$ it reduces
to $d\lesssim2\lfloor1/R\rfloor$.

The factor graph representation of a linear code serves as the foundation
for a belief propagation decoder. Instead of determining the most
likely error as given in Eq. (\ref{eq:syndrome decoder}) a belief
propagation decoder approximates it in a symbol-wise fashion. This
gives an estimate $\hat{\boldsymbol{e}}=(\hat{e}_{1},\dots,\hat{e}_{n})$
where
\begin{equation}
\hat{e}_{j}=\underset{e_{j}\in\mathrm{GF}(q)}{\mathrm{argmax}}P(e_{j}|\boldsymbol{z}).\label{eq:symbol prob}
\end{equation}
An expression for $P(e_{j}|\boldsymbol{z})$ can be obtained by marginalizing
$P(\boldsymbol{e}|\boldsymbol{z})$. Assuming that the error components
are independent 
\begin{align}
P(\boldsymbol{e}|\boldsymbol{z})\propto & \prod_{l=1}^{n}P(e_{l})\delta(H\boldsymbol{e}=\boldsymbol{z})\nonumber \\
= & \prod_{l=1}^{n}P(e_{l})\prod_{i=1}^{m}\delta\left(\sum_{j=1}^{n}H_{ij}e_{j}=z_{i}\right).
\end{align}
Fixing $e_{j}=a$ and summing over all other components gives

\begin{equation}
P(e_{j}=a|\boldsymbol{z})\propto\sum_{\boldsymbol{e}:e_{j}=a}\prod_{l=1}^{n}P(e_{l})\prod_{i=1}^{m}\delta\left(\sum_{j=1}^{n}H_{ij}e_{j}=z_{i}\right).\label{eq:symbol wise prob}
\end{equation}
Belief propagation efficiently approximates these marginals by passing
messages on the code's factor graph. For a code over $\mathrm{GF}(q)$,
these messages will be vectors of length $q$. Initially, a message
is sent from every error node $v_{j}$ to the check nodes in the neighborhood
$\mathcal{N}(v_{j})=\{c_{i}\in C:\{c_{i},v_{j}\}\in E\}$. In particular,
the message sent to check node $c_{i}\in\mathcal{N}(v_{j})$ is $\mu_{j\rightarrow i}$
where the element corresponding to $a\in\mathrm{GF}(q)$ is
\begin{equation}
\mu_{j\rightarrow i}^{a}=P(e_{j}=a).\label{eq:classical bp initial message}
\end{equation}
Note that this message simply gives the channel error probabilities.
Every check node $c_{i}$ then sends a message back to the error nodes
in the neighborhood $\mathcal{M}(c_{i})=\{v_{j}\in V:\{c_{i},v_{j}\}\in E\}$.
In particular, the message sent to error node $v_{j}\in\mathcal{M}(c_{i})$
is $\lambda_{i\rightarrow j}$ with
\begin{equation}
\lambda_{i\rightarrow j}^{a}=K\sum_{\boldsymbol{e}:e_{j}=a}\delta\left(\sum_{j'\in\mathcal{M}(i)}H_{ij'}e_{j'}=z_{i}\right)\prod_{j'\in\mathcal{M}(i)\backslash j}\mu_{j'\rightarrow i}^{e_{j'}},\label{eq:check to error message}
\end{equation}
where, through slight abuse of notation, $\mathcal{M}(i)=\{j\in\{1,\dots,n\}:v_{j}\in\mathcal{M}(c_{i})\}$
and $K$ is a normalization factor chosen such that $\sum_{a}\lambda_{i\rightarrow j}^{a}=1$.
An estimate of the marginal probability $P(e_{j}|\boldsymbol{z})$
can then be made with 
\begin{equation}
\hat{P}(e_{j}=a|\boldsymbol{z})=KP(e_{j}=a)\prod_{i\in\mathcal{N}(j)}\lambda_{i\rightarrow j}^{a},\label{eq:marignal probability estimate}
\end{equation}
where $\mathcal{N}(j)=\{i\in\{1,\dots,m\}:c_{i}\in\mathcal{N}(v_{j})\}$
and $K$ is a normalization factor. From this $\hat{\boldsymbol{e}}$
can be estimated in a symbol-wise fashion as in Eq. (\ref{eq:symbol prob}).
If $\hat{\boldsymbol{z}}=H\hat{\boldsymbol{e}}=\boldsymbol{z}$, then
decoding is complete; otherwise, another message is sent from each
error node to its connected check nodes. The elements of this message
are
\begin{equation}
\mu_{j\rightarrow i}^{a}=KP(e_{j}=a)\prod_{i'\in\mathcal{N}(j)\backslash i}\lambda_{i'\rightarrow j}^{a},\label{eq:error to check message}
\end{equation}
where $K$ is again a normalization factor. There is then another
round of check to error node messages as in Eq. (\ref{eq:check to error message}),
followed by an approximation of marginals as in Eq. (\ref{eq:marignal probability estimate}).
This process of sending error to check messages followed by check
to error messages and a computation of marginals proceeds iteratively
until either $\hat{\boldsymbol{z}}=\boldsymbol{z}$ or a maximum number
of iterations $I_{\max}$ is reached. The most computationally complex
component of belief propagation is the check to error node message
calculation of Eq. (\ref{eq:check to error message}). However, it
can be performed efficiently using a Fourier transform as outlined
in Appendix \ref{sec:Belief-propagation-Fourier}. 

There are two types of decoding error exhibited by a belief propagation
decoder. The first type is the detected error where decoding ends
with $\hat{\boldsymbol{z}}\neq\boldsymbol{z}$ (and hence $\hat{\boldsymbol{e}}\neq\boldsymbol{e}$).
Such errors do not occur when using an optimal decoder and as such
are fundamentally a failing of the belief propagation decoder itself.
The second type of error is the undetected error where decoding ends
with $\hat{\boldsymbol{z}}=\boldsymbol{z}$ but $\hat{\boldsymbol{e}}\neq\boldsymbol{e}$.
These are the same type of error exhibited by the optimal decoder
and as such can be attributed to a failing of the code. It therefore
follows that for a symmetric channel using a code with a lower distance
will tend to result in a higher rate of undetected errors.

Belief propagation decoding is an approximation on two levels. Firstly,
it assumes that the most likely error is equal to the symbol-wise
most likely error. Secondly, the estimate of the symbol-wise most
likely error is based on the approximate marginal probabilities $\hat{P}(e_{j}|\boldsymbol{z})$
that are only exact when the code's factor graph is a tree \citep{richardson2008modern},
which as previously outlined is unlikely. However, good decoding performance
can still be achieved when the factor graph is sparsely connected
\citep{richardson2008modern}. Linear codes with such a representation
are called low-density parity-check (LDPC) codes (most codes do not
have such a representation \citep{mackay2003information,richardson2008modern}).
Decoding performance is further improved when the factor graph contains
few short cycles \citep{mcgowan2003loop}.

\subsection{Stabilizer codes\label{subsec:Stabilizer-codes}}

The action of a quantum channel $\Phi$ on a quantum state described
by the density operator $\rho$ is
\begin{equation}
\Phi(\rho)=\sum_{k}A_{k}\rho A_{k}^{\dagger},
\end{equation}
where the $A_{k}$, called Kraus operators, satisfy $\sum_{k}A_{k}^{\dagger}A_{k}=I$
(the identity operator) \citep{kraus1983states}. In this paper we
are interested in qubit systems; that is, systems where states $|\phi\rangle$
belong to a two dimensional Hilbert space $\mathcal{H}\cong\mathbb{C}^{2}$.
Furthermore, we are concerned with Pauli channels. These are channels
of the form
\begin{equation}
\Phi(\rho)=p_{I}\rho+p_{X}X\rho X+p_{Y}Y\rho Y+p_{Z}Z\rho Z,
\end{equation}
where in the computational $\{|0\rangle,|1\rangle\}$ basis
\begin{equation}
X=\left(\begin{array}{cc}
0 & 1\\
1 & 0
\end{array}\right),\,Y=\left(\begin{array}{cc}
0 & -i\\
i & 0
\end{array}\right),\,Z=\left(\begin{array}{cc}
1 & 0\\
0 & -1
\end{array}\right).
\end{equation}
The action of this channel can be interpreted as mapping a pure state
$|\phi\rangle$ to $E|\phi\rangle$ where the error $E$ is $I$ with
probability $p_{I}$, $X$ with probability $p_{X}$, $Y$ with probability
$p_{Y}$, or $Z$ with probability $p_{Z}$ \citep{nielsen2002quantum}.
$X$ can be viewed as a bit flip operator as $X|0\rangle=|1\rangle$
and $X|1\rangle=|0\rangle$. $Z$ can be viewed as a phase flip as
$Z|0\rangle=|0\rangle$ and $Z|1\rangle=-|1\rangle$. $Y=iXZ$ can
be viewed as a combined bit and phase flip. Of particular interest
is the depolarizing channel where $p_{I}=1-p$ and $p_{X}=p_{Y}=p_{Z}=p/3$.
We are also interested in the $XZ$ channel for which the $X$ and
$Z$ components of an error $E\propto X^{u}Z^{v}$, where $u,v\in\mathrm{GF}(2)$,
occur independently with equal probability $q$. It follows from the
independence of the error components that $p_{X}=p_{Z}=q(1-q)$ and
$p_{Y}=q^{2}$. These values can be expressed in terms of the total
error probability $p=p_{X}+p_{Y}+p_{Z}$ as $q=1-\sqrt{1-p}$, $p_{X}=p_{Z}=\sqrt{1-p}(1-\sqrt{1-p})$,
and $p_{Y}=(1-\sqrt{1-p})^{2}$.

The Pauli matrices are Hermitian, unitary, and anticommute with each
other. Furthermore, they form a group called the Pauli group
\begin{equation}
\mathcal{P}_{1}=\{\pm I,\pm iI,\pm X,\pm iX,\text{\ensuremath{\pm Y,\pm iY,\pm Z,\pm iZ}\}=\ensuremath{\langle X,Y,Z\rangle}.}
\end{equation}
The $n$-qubit Pauli group $\mathcal{P}_{n}$ is defined as all $n$-fold
tensor product combinations of elements of $\mathcal{P}_{1}$. For
example, $\mathcal{P}_{8}$ contains the element $I\otimes I\otimes X\otimes I\otimes Y\otimes Z\otimes I\otimes I$,
which is often written more compactly as $IIXIYZII$ or $X_{3}Y_{5}Z_{6}$.
The weight of some $g\in\mathcal{P}_{n}$ is the number of elements
in the tensor product that are not equal to the identity up to phase.
The commutation relations of the Pauli matrices mean that elements
of $\mathcal{P}_{n}$ must either commute or anticommute, with two
elements anticommuting if their non-identity components differ in
an odd number of places. 

Similar to the classical case, the noise introduced by a quantum channel
can be protected against by employing a code. A quantum (qubit) code
is a subspace $\mathcal{Q}\subseteq(\mathbb{C}^{2})^{\otimes n}$.
Codewords $|\phi\rangle\in\mathcal{Q}$ are transmitted across the
combined $n$-qubit channel $\Phi^{\otimes n}$. If $\Phi$ is a Pauli
channel, then $\Phi^{\otimes n}$ maps codewords $|\phi\rangle$ to
$E|\phi\rangle$ where $E\in\mathcal{P}_{n}$. Assuming the channel
acts on each qubit independently, the probability of an error $E$
occurring (up to phase) is
\begin{equation}
P(E)=\prod_{i=1}^{n}P(E_{i}),
\end{equation}
where $P(E_{i})$ is the probability of the error $E_{i}$ occurring
(up to phase) on the single qubit channel $\Phi$. Note that errors
are considered up to phase as the resulting state is equivalent up
to such a phase factor. A convenient way of handling this is to group
errors in $\mathcal{P}_{n}$ up to phase with $\tilde{E}=\{E,-E,iE,-iE\}\in\mathcal{P}_{n}/\{\pm I,\pm iI\}=\tilde{\mathcal{P}}_{n}$.

Stabilizer codes are defined by an abelian subgroup $\mathcal{S}<\mathcal{P}_{n}$,
called the stabilizer, that does not contain $-I$ \citep{gottesman1997stabilizer}.
The code $\mathcal{Q}$ is the space of states that are fixed by every
element $s_{i}\in\mathcal{S}$; that is,
\begin{equation}
\mathcal{Q}=\{|\phi\rangle\in(\mathbb{C}^{2})^{\otimes n}:s_{i}|\phi\rangle=|\phi\rangle\,\forall\,s_{i}\in\mathcal{S}\}.
\end{equation}
The requirement that $-I\notin\mathcal{S}$ both means that no $s\in\mathcal{S}$
can have a phase factor of $\pm i$, and that if $s\in\mathcal{S}$,
then $-s\notin\mathcal{S}$. If $\mathcal{S}$ is generated by $M=\{M_{1},\dots,M_{m}\}\subset\mathcal{P}_{n}$,
then it is sufficient (and obviously necessary) for $\mathcal{Q}$
to be stabilized by every $M_{i}$. Assuming that the set of generators
is minimal, it can be shown that $\dim(\mathcal{Q})=2^{n-m}=2^{k}$
\citep{nielsen2002quantum}; that is, $\mathcal{Q}$ encodes the state
of a $k$-qubit system. If the generators of $\mathcal{S}$ are sparse,
then $\mathcal{Q}$ is called a quantum LDPC (QLDPC) code.

Suppose an error $E$ occurs mapping some codeword $|\phi\rangle\in\mathcal{\mathcal{Q}}$
to $E|\phi\rangle$. A projective measurement of a generator $M_{i}$
will give the result $+1$ if $[E,M_{i}]=EM_{i}-M_{i}E=0$ or $-1$
if $\{E,M_{i}\}=EM_{i}+M_{i}E=0$. These measurement values define
the syndrome $\boldsymbol{z}\in\mathrm{GF}(2)^{m}$ with 
\begin{equation}
z_{i}=\begin{cases}
0 & \mathrm{if}\,[E,M_{i}]=0,\\
1 & \mathrm{if}\,\{E,M_{i}\}=0.
\end{cases}
\end{equation}
There are three classes of error that can occur. The first class are
those errors $\tilde{E}=\{E,-E,iE,-iE\}\in\tilde{\mathcal{S}}$ where
$\tilde{\mathcal{S}}$ is the group
\begin{equation}
\tilde{\mathcal{S}}=\{\tilde{s}=\{s,-s,is,-is\}:s\in\mathcal{S}\}.\label{eq:stabilizer up to phase}
\end{equation}
Such errors have no effect on the code and result in the trivial syndrome
$\boldsymbol{z}=\boldsymbol{0}$ (as the stabilizer is abelian). The
second class of errors are those $\tilde{E}\in C(\tilde{\mathcal{S}})\backslash\tilde{\mathcal{S}}$
where $C(\tilde{S})$ is the centralizer of $\tilde{\mathcal{S}}$
in $\tilde{\mathcal{P}}_{n}$, which in this case is actually equal
to $N(\tilde{\mathcal{S}})$ (the normalizer of $\tilde{\mathcal{S}}$
in $\tilde{\mathcal{P}}_{n}$) \citep{gottesman1997stabilizer}. These
are errors that commute with every stabilizer and therefore also yield
$\boldsymbol{z}=\boldsymbol{0}$; however, the effect of such errors
on the code is non-trivial. The final class of errors are those $\tilde{E}\in\tilde{\mathcal{P}}_{n}\backslash N(\tilde{\mathcal{S}})$,
which yield non-trivial syndromes $\boldsymbol{z}\neq\boldsymbol{0}$
and also act non-trivially on the code. In general, the syndrome resulting
from some error $\tilde{E}\in\tilde{\mathcal{P}}_{n}$ depends only
on which coset of $\tilde{\mathcal{P}}_{n}/N(\tilde{\mathcal{S}})$
it belongs to, while its effect on the code depends only on which
coset of $\tilde{\mathcal{P}}_{n}/\tilde{\mathcal{S}}$ it belongs
to (note that $\tilde{\mathcal{S}}\vartriangleleft N(\tilde{\mathcal{S}})\vartriangleleft\tilde{\mathcal{P}}_{n}$
as $\tilde{\mathcal{P}}_{n}$ is abelian). This phenomena of distinct
errors having an identical effect on a code is called degeneracy and
has no classical analog. In the classical case, the distance $d$
of a linear code is equal to the weight of the lowest weight error
yielding a trivial syndrome while having a non-trivial effect on the
code. This extends to the quantum case, with the distance $d$ of
a stabilizer code being the weight of the lowest weight element in
$N(\tilde{\mathcal{S}})\backslash\tilde{\mathcal{S}}$ \citep{gottesman1997stabilizer}.
An $n$-qubit code of dimension $2^{k}$ with distance $d$ is called
an $[[n,k]]$ or $[[n,k,d]]$ code (the double brackets differentiate
it from a classical code).

From a decoding point of view, the syndrome measurement determines
which coset of $\tilde{\mathcal{P}}_{n}/N(\tilde{\mathcal{S}})$ an
error $\tilde{E}$ belongs to. If this coset has the representative
$\tilde{g}\in\tilde{\mathcal{P}}_{n}$, then an ideal decoder determines
the coset $\hat{A}$ in $(\tilde{g}N(\tilde{\mathcal{S}}))/\tilde{\mathcal{S}}$
that $\tilde{E}$ is most likely to belong to. Importantly, $\hat{A}$
does not necessarily contain the individually most likely error in
$\tilde{g}N(\tilde{\mathcal{S}})$. If $\hat{A}$ has the representative
$\tilde{\hat{E}}=\{\hat{E},-\hat{E},i\hat{E},-i\hat{E}\}$, then the
decoder attempts to correct the channel error by applying $\hat{E}$
to the channel output. If $\tilde{E}\in\hat{A}$, then $\tilde{\hat{E}}\tilde{E}\in\tilde{\mathcal{S}}$
and as such this process corrects the error; otherwise, if $\tilde{E}\notin\hat{A}$,
then a decoding error has occurred. Similar to the classical case,
the probability of a decoding failure given some syndrome measurement
$\boldsymbol{z}$ is 
\begin{equation}
P(\tilde{E}\notin\hat{A}|\boldsymbol{z})=\frac{P(\tilde{g}N(\tilde{\mathcal{S}}))-P(\hat{A})}{P(\tilde{g}N(\tilde{\mathcal{S}}))},
\end{equation}
where $P(\tilde{g}N(\tilde{\mathcal{S}}))$ and $P(\hat{A})$ are
the probabilities of an error being in $\tilde{g}N(\tilde{\mathcal{S}})$
or $\hat{A}$ respectively. From this it follows that the probability
of a decoding error is high if the probability distribution over $(\tilde{g}N(\tilde{\mathcal{S}}))/\tilde{\mathcal{S}}$
is not sharply peaked, which will occur if $N(\tilde{\mathcal{S}})\backslash\tilde{\mathcal{S}}$
contains high probability errors. For the depolarizing channel this
corresponds to $N(\tilde{\mathcal{S}})\backslash\tilde{\mathcal{S}}$
containing low weight elements, meaning that the distance $d$ gives
some indication of decoder performance.

\subsection{Stabilizer code representations\label{subsec:Stabilizer-code-representations}}

It is possible to represent elements of $\tilde{\mathcal{P}}_{1}$
as elements of $\mathrm{GF}(2)^{2}$ according to the isomorphism
\citep{calderbank1997quantum,gottesman1997stabilizer} 
\begin{equation}
I\leftrightarrow(0,0),\,X\leftrightarrow(1,0),\,Y\propto XZ\leftrightarrow(1,1),\,Z\leftrightarrow(0,1).
\end{equation}
This can be extended to elements of $\tilde{\mathcal{P}}_{n}$ according
to
\begin{equation}
X^{u_{1}}Z^{v_{1}}\otimes\dots\otimes X^{u_{n}}Z^{v_{n}}\leftrightarrow(u_{1},\dots,u_{n}|v_{1},\dots,v_{n}).
\end{equation}
This can be written more compactly as $X^{\boldsymbol{u}}Z^{\boldsymbol{v}}\leftrightarrow(\boldsymbol{u}|\boldsymbol{v})\in\mathrm{GF}(2)^{2n}$
where $\boldsymbol{u}=(u_{1},\dots,u_{n}),\boldsymbol{v}=(v_{1},\dots,v_{n})\in\mathrm{GF}(2)^{n}$.
The product of elements in $\tilde{\mathcal{P}}_{n}$ corresponds
to addition in $\mathrm{GF}(2)^{2n}$. Representatives of elements
in $\tilde{\mathcal{P}}_{n}$ commute if the symplectic inner product
of the binary representations is zero; otherwise, they anticommute.
Note that the symplectic inner product of $\boldsymbol{a}=(\boldsymbol{u}|\boldsymbol{v})\in\mathrm{GF}(2)^{2n}$
and $\boldsymbol{b}=(\boldsymbol{u}'|\boldsymbol{v}')\in\mathrm{GF}(2)^{2n}$
is
\begin{equation}
\boldsymbol{a}\circ\boldsymbol{b}=\boldsymbol{u}\cdot\boldsymbol{v}'+\boldsymbol{u}'\cdot\boldsymbol{v}=\sum_{i=1}^{n}(u_{i}v'_{i}+u'_{i}v_{i}).
\end{equation}
Considering $\boldsymbol{a}$ and $\boldsymbol{b}$ as row vectors,
this simplifies to $\boldsymbol{a}\circ\boldsymbol{b}=\boldsymbol{a}P\boldsymbol{b}^{T}$
where $P$ is the $2n\times2n$ matrix
\begin{equation}
P=\left(\begin{array}{cc}
0 & I\\
I & 0
\end{array}\right).
\end{equation}

The binary representations of the $m$ generators of some stabilizer
$\mathcal{S}$ define the rows of an $m\times2n$ binary matrix $H$.
This matrix has the form
\begin{equation}
H=(H_{X}|H_{Z}),\label{eq:gf2 pcm}
\end{equation}
where $H_{X}$ and $H_{Z}$ are each $m\times n$ matrices. Note that
while $H$ only defines a stabilizer up to phase $\tilde{\mathcal{S}}$,
the codes defined by different stabilizers corresponding to $\tilde{\mathcal{S}}$
will all have the same error correction properties. Considering $H$
as the parity-check matrix of a classical binary code $\mathcal{C}$,
the stabilizer elements correspond to elements of the dual code $\mathcal{C}^{\perp}$.
The requirement that all stabilizer generators commute becomes 
\begin{equation}
H_{X}H_{Z}^{T}+H_{Z}H_{X}^{T}=0.\label{eq:gf2 comm relation}
\end{equation}
Any classical linear code with a parity-check matrix $H$ that satisfies
this constraint can be used to define a stabilizer code. Furthermore,
if $H$ is sparse, then this stabilizer code is a QLDPC code. Errors
can also be considered within the binary framework. Suppose that some
error $E\propto X^{\boldsymbol{e}_{X}}Z^{\boldsymbol{e}_{Z}}$ occurs.
This error has the binary representation $\boldsymbol{e}=(\boldsymbol{e}_{X}^{T}|\boldsymbol{e}_{Z}^{T})^{T}$,
and the corresponding syndrome is simply $\boldsymbol{z}=HP\boldsymbol{e}$
(where $\boldsymbol{e}_{X}$, $\boldsymbol{e}_{Z}$, and $\boldsymbol{e}$
are column vectors for consistency with the classical case).

A subclass of stabilizer codes are the Calderbank-Shor-Steane (CSS)
codes \citep{calderbank1996good,steane1996multiple}, which have a
binary representation of the form
\begin{equation}
H=\left(\begin{array}{c|c}
\tilde{H}_{X} & 0\\
0 & \tilde{H}_{Z}
\end{array}\right).\label{eq:css pcm}
\end{equation}
The commutation condition of Eq. (\ref{eq:gf2 comm relation}) becomes
$\tilde{H}_{Z}\tilde{H}_{X}^{T}=0$ (or equivalently $\tilde{H}_{X}\tilde{H}_{Z}^{T}=0$).
Considering $\tilde{H}_{X}$ and $\tilde{H}_{Z}$ as parity-check
matrices for classical codes $\mathcal{C}_{X}$ and $\mathcal{C}_{Z}$
respectively this commutation condition requires that $\mathcal{C}_{X}^{\perp}\subseteq\mathcal{C}_{Z}$
(or equivalently $\mathcal{C}_{Z}^{\perp}\subseteq\mathcal{C}_{X}$).
If $\tilde{H}_{Z}=\tilde{H}_{X}$, then $\mathcal{C}_{Z}=\mathcal{C}_{X}$,
which gives $\mathcal{C}_{X}^{\perp}\subseteq\mathcal{C}_{X}$. Such
codes are called dual-containing CSS codes.

Elements of $\tilde{\mathcal{P}}_{1}$ can also be represented as
elements of $\mathrm{GF}(4)=\{0,1,\omega,\omega^{2}=\bar{\omega}\}$
according to the isomorphism \citep{calderbank1996quantum,gottesman1997stabilizer}
\begin{equation}
I\leftrightarrow0,\,X\leftrightarrow1,\,Y\leftrightarrow\bar{\omega},\,Z\leftrightarrow\omega.
\end{equation}
Elements of $\tilde{\mathcal{P}}_{n}$ then map to elements of $\mathrm{GF}(4)^{n}$,
with the product of elements in $\tilde{\mathcal{P}}_{n}$ corresponding
to addition in $\mathrm{GF}(4)^{n}$ ($\mathrm{GF}(4)$ addition and
multiplication are defined in Tables \ref{gf4 add} and \ref{gf4 mult}
respectively). Representatives of elements in $\tilde{\mathcal{P}}_{n}$
commute if the trace inner product of the corresponding elements of
$\mathrm{GF}(4)^{n}$ is zero. Note that the trace inner product of
$\boldsymbol{a},\boldsymbol{b}\in\mathrm{GF}(4)^{n}$ is
\begin{equation}
\boldsymbol{a}*\boldsymbol{b}=\mathrm{tr}(\boldsymbol{a}\cdot\bar{\boldsymbol{b}})=\mathrm{tr}\left(\sum_{i=1}^{n}a_{i}\bar{b_{i}}\right),
\end{equation}
where $\bar{0}=0$, $\bar{1}=1$, $\bar{\omega}=\omega^{2}$, and
$\bar{\omega^{2}}=\omega$; and $\mathrm{tr}(x)=x+\bar{x}$ (that
is, $\mathrm{tr}(0)=\mathrm{tr}(1)=0$ and $\mathrm{tr}(\omega)=\mathrm{tr}(\bar{\omega})=1$).

\begin{table}
\caption{\label{gf4 add} $\mathrm{GF}(4)$ addition.}
\begin{tabular}{c|cccc}
$+$ & $0$ & $1$ & $\omega$ & $\bar{\omega}$\tabularnewline
\hline 
$0$ & $0$ & $1$ & $\omega$ & $\bar{\omega}$\tabularnewline
$1$ & $1$ & $0$ & $\bar{\omega}$ & $\omega$\tabularnewline
$\omega$ & $\omega$ & $\bar{\omega}$ & $0$ & $1$\tabularnewline
$\bar{\omega}$ & $\bar{\omega}$ & $\omega$ & $1$ & $0$\tabularnewline
\end{tabular}
\end{table}

\begin{table}
\caption{\label{gf4 mult}$\mathrm{GF}(4)$ multiplication.}
\begin{tabular}{c|cccc}
$\times$ & $0$ & $1$ & $\omega$ & $\bar{\omega}$\tabularnewline
\hline 
$0$ & $0$ & $0$ & $0$ & $0$\tabularnewline
$1$ & $0$ & $1$ & $\omega$ & $\bar{\omega}$\tabularnewline
$\omega$ & $0$ & $\omega$ & $\bar{\omega}$ & $1$\tabularnewline
$\bar{\omega}$ & $0$ & $\bar{\omega}$ & $1$ & $\omega$\tabularnewline
\end{tabular}
\end{table}

The $\mathrm{GF}(4)^{n}$ representations of the $m$ generators of
some stabilizer $\mathcal{S}$ define an $m\times n$ $\mathrm{GF}(4)$
matrix $H$ in much the same way as the binary case. A stabilizer
with the $\mathrm{GF}(2)$ representation of Eq. (\ref{eq:gf2 pcm})
has the $\mathrm{GF}(4)$ representation
\begin{equation}
H=H_{X}+\omega H_{Z}.\label{eq:gf4 pcm}
\end{equation}
For a CSS code this becomes
\begin{equation}
H=\left(\begin{array}{c}
\tilde{H}_{X}\\
\omega\tilde{H}_{Z}
\end{array}\right),\label{eq:gf4 dual containing css pcm}
\end{equation}
with $\tilde{H}_{X}$ and $\tilde{H}_{Z}$ as defined in Eq. (\ref{eq:css pcm}).
The stabilizer corresponds to the additive group generated by the
rows of $H$. This group can be considered as an additive classical
code $\mathcal{C}$ over $\mathrm{GF}(4)$. The rows of $H$ must
be orthogonal with respect to the trace inner product. Therefore,
if $\mathcal{C}^{\perp}$ is the dual code of $\mathcal{C}$ with
respect to the trace inner product, then $\mathcal{C}\subseteq\mathcal{C}^{\perp}$.
Any such self-orthogonal additive $\mathrm{GF}(4)$ code can be used
to define a stabilizer code. Errors can also be considered in the
$\mathrm{GF}(4)$ framework. An error $E$ with $\mathrm{GF}(4)$
representation $\boldsymbol{e}$ (again, taken to be a column vector)
will yield a syndrome $\boldsymbol{z}=\mathrm{tr}(H\boldsymbol{e})$.
Note that while $H$ is a generator matrix for $\mathcal{C}$, we
essentially consider it as a parity-check matrix because of the role
it plays in syndrome calculation and hence in belief propagation decoding.

\subsection{Belief propagation decoding for stabilizer codes}

Belief propagation decoding can be applied to stabilizer codes using
the $\mathrm{GF}(2)$ and $\mathrm{GF}(4)$ representations of the
previous section. Such a belief propagation decoder aims to estimate
the symbol-wise most likely error (up to phase) $\hat{E}=\hat{E}_{1}\otimes\dots\otimes\hat{E}_{n}$
where
\begin{equation}
\hat{E}_{j}=\underset{E_{j}}{\mathrm{argmax}}\,P(E_{j}|\boldsymbol{z}).
\end{equation}
A $\mathrm{GF}(4)$ based belief propagation decoder can be used for
any QLDPC code. This decoder attempts to make a symbol-wise estimate
$\hat{\boldsymbol{e}}\in\mathrm{GF}(4)^{n}$ that maps to $\hat{E}$
according to the isomorphism outlined in Sec. \ref{subsec:Stabilizer-code-representations}.
The $\mathrm{GF}(4)$ decoder behaves very similarly to the belief
propagation decoder presented for classical linear codes in Section
\ref{subsec:LDPC-codes-and}. The only change is to account for the
difference in syndrome calculation. In particular, the check to error
node message is modified to
\begin{align}
\lambda_{i\rightarrow j}^{a}= & K\sum_{\boldsymbol{e}:e_{j}=a}\delta\left(\mathrm{tr}(\sum_{j'\in\mathcal{M}(i)}H_{ij'}\bar{e_{j'}})=z_{i}\right)\nonumber \\
 & \times\prod_{j'\in\mathcal{M}(i)\backslash j}\mu_{j'\rightarrow i}^{e_{j'}}.\label{eq:quantum check node}
\end{align}
This calculation can also be performed efficiently using a Fourier
transform as outlined in Appendix \ref{sec:Stabiliser-code-Fourier}.
The channel error probabilities used in error to check node messages
(Eqs. (\ref{eq:classical bp initial message}) and (\ref{eq:error to check message}))
and in marginal calculation (Eq. (\ref{eq:marignal probability estimate}))
are $P(e_{j}=0)=1-p$, $P(e_{j}=1)=p_{X}$, $P(e_{j}=\bar{\omega})=p_{Y}$,
and $P(e_{j}=\omega)=p_{Z}$.

For the subclass of CSS codes it is also possible to use two separate
$\mathrm{GF}(2)$ based belief propagation decoders. For some error
$E\propto X^{\boldsymbol{e}_{X}}Z^{\boldsymbol{e}_{Z}}$ the corresponding
binary error is $\boldsymbol{e}=(\boldsymbol{e}_{X}^{T}|\boldsymbol{e}_{Z}^{T})^{T}$
, which yields the syndrome
\begin{equation}
\boldsymbol{z}=HP\boldsymbol{e}=\left(\begin{array}{c}
\tilde{H}_{X}\boldsymbol{e}_{Z}\\
\tilde{H}_{Z}\boldsymbol{e}_{X}
\end{array}\right)=\left(\begin{array}{c}
\boldsymbol{z}_{Z}\\
\boldsymbol{z}_{X}
\end{array}\right).
\end{equation}
Using $\boldsymbol{z}_{Z}$ and $\tilde{H}_{X}$ an estimate $\hat{\boldsymbol{e}}_{Z}$
of $\boldsymbol{e}_{Z}$ can be made using a classical binary belief
propagation decoder. The same can be done with $\boldsymbol{z}_{X}$
and $\tilde{H}_{Z}$ to make an estimate $\hat{\boldsymbol{e}}_{X}$
of $\boldsymbol{e}_{X}$. The $j$-th component of $\boldsymbol{e}_{X}$,
denoted $e_{X}^{(j)}$, is equal to one if $E_{j}\propto X$ or $E_{j}\propto Y$.
Therefore, $P(e_{X}^{(j)}=1)=p_{X}+p_{Y}$ and similarly $P(e_{Z}^{(j)}=1)=p_{Y}+p_{Z}$.
These values are used as the channel error probabilities for the two
decoders, which amounts to considering the quantum channel as two
binary symmetric channels. Note that for depolarizing channel $P(e_{X}^{(j)}=1)=P(e_{Z}^{(j)}=1)=2p/3$,
while for the $XZ$ channel $P(e_{X}^{(j)}=1)=P(e_{Z}^{(j)}=1)=1-\sqrt{1-p}$.

As in the classical case, belief propagation decoding can result in
both detected and undetected errors. If $\hat{\boldsymbol{z}}\neq\boldsymbol{z}$,
where $\hat{\boldsymbol{z}}$ is the syndrome associated with the
error estimate $\hat{E}$, then a detected error has occurred. Again,
these detected errors are a failing of the decoder. If $\hat{\boldsymbol{z}}=\boldsymbol{z}$
but $\tilde{\hat{E}}\tilde{E}\notin\tilde{\mathcal{S}}$, then an
undetected error has occurred, which is fundamentally a failing of
the code itself. It therefore follows that for the depolarizing channel,
using a code with a lower distance will tend to result in a higher
rate of undetected errors.

Using belief propagation in the quantum case is an even greater approximation
than in the classical case. As outlined in Sec. \ref{subsec:Stabilizer-codes},
an optimal decoder for a stabilizer code will determine the most likely
coset of errors rather than the single most likely error. By definition,
QLDPC codes have many low weight stabilizers, which means there will
be a large number of elements of the most likely coset with similar
weight and hence similar probability. This spreading of probability
increases the chance that the single most likely error will not belong
to the most likely coset of errors. Approximating the ideal decoder
with one that determines the single most likely error will therefore
lead to an increased error rate. Belief propagation goes one step
further away from the optimal decoder by estimating the single most
likely error in a symbol-wise fashion, which can lead to so-called
symmetric degeneracy errors. Such errors are well explained by the
example of Ref. \citep{poulin2008iterative}, which is as follows.
Consider a two-qubit stabilizer code with generators $M_{1}=XX$ and
$M_{2}=ZZ$, and assume that the error $E=IX$ occurs leading to a
syndrome $\boldsymbol{z}=(0,1)^{T}$. The coset of errors that give
this syndrome is $\{XI,IX,YZ,ZY\}$ (grouping errors up to phase).
As a result, the error probabilities on both qubits are $P(E_{i}=I|z)=P(E_{i}=X|z)=Kp_{I}p_{X}$
and $P(E_{i}=Y|z)=P(E_{i}=Z|z)=Kp_{Y}p_{Z}$ where $K=1/(2p_{I}p_{X}+2p_{Y}p_{Z})$.
This symmetry of error probabilities results in the decoder estimating
the same error on each qubit. This is not a symmetry exhibited by
any of the errors that yield $\boldsymbol{z}$ and as such even an
ideal symbol-wise decoder will yield a detected error.

The requirement that all stabilizer generators must commute also degrades
belief propagation performance as it results in $4$-cycles. Consider
some qubit $j$, there must be (at least) two stabilizer generators,
say $M_{i}$ and $M_{i'}$, that act non-trivially on $j$ with different
Pauli matrices. If this is not the case, then there will be a weight
one element of $N(\tilde{\mathcal{S}})\backslash\tilde{\mathcal{S}}$,
meaning that the code will have distance $d=1$ (making it of little
to no interest). As $M_{i}$ and $M_{i'}$ contain different Pauli
matrices in position $j$, they must also contain different Pauli
matrices at some other position $j'$ to ensure that they commute
with each other. This results in a $4$-cycle in the $\mathrm{GF}(4)$
factor graph as check nodes $c_{i}$ and $c_{i'}$ both connect to
error nodes $v_{j}$ and $v_{j'}$. In the case of a CSS code, any
$4$-cycles resulting from an overlap between one row from $\tilde{H}_{X}$
and one row from $\tilde{H}_{Z}$ can be removed by decoding with
a pair of $\mathrm{GF}(2)$ decoders rather than a $\mathrm{GF}(4)$
decoder. If it is a dual-containing CSS code, then there must still
be $4$-cycles in the $\mathrm{GF}(2)$ factor graph as the rows of
$\tilde{H}=\tilde{H}_{X}=\tilde{H}_{Z}$ must overlap in an even number
of positions to ensure that $\tilde{H}\tilde{H}^{T}=0$. If the code
is non-dual-containing, then it is possible for $\tilde{H}_{X}$ and
$\tilde{H}_{Z}$ to have corresponding $\mathrm{GF}(2)$ factor graphs
with no $4$-cycles.

The reduction in $4$-cycles, along with the reduced inherent complexity,
makes $\mathrm{GF}(2)$ decoding attractive for CSS codes. However,
treating a Pauli channel as a pair of binary symmetric channels ignores
potential correlations between the $X$ and $Z$ components of an
error $E\propto X^{\boldsymbol{e}_{X}}Z^{\boldsymbol{e}_{Z}}$. These
correlations are described by the conditional probabilities
\begin{align}
P(e_{Z}^{(j)}=1|e_{X}^{(j)}=1)= & \frac{p_{Y}}{p_{X}+p_{Y}},\label{eq:cond prob 1}\\
P(e_{Z}^{(j)}=1|e_{X}^{(j)}=0)= & \frac{p_{Z}}{1-(p_{X}+p_{Y})},\\
P(e_{X}^{(j)}=1|e_{Z}^{(j)}=1)= & \frac{p_{Y}}{p_{Y}+p_{Z}},\\
P(e_{X}^{(j)}=1|e_{Z}^{(j)}=0)= & \frac{p_{X}}{1-(p_{Y}+p_{Z})}.\label{eq:cond prob 4}
\end{align}
The $X$ and $Z$ components are uncorrelated if they occur independently,
which requires $P(e_{Z}^{(j)}=1|e_{X}^{(j)})=P(e_{Z}^{(j)}=1)=p_{Y}+p_{Z}$
and $P(e_{X}^{(j)}=1|e_{Z}^{(j)})=P(e_{X}^{(j)}=1)=p_{X}+p_{Y}$.
This is equivalent to the requirement that $p_{Y}=(p_{X}+p_{Y})(p_{Y}+p_{Z})$,
which is satisfied by the $XZ$ channel but not by the depolarizing
channel.

\section{Modified decoders\label{sec:Modified-decoders}}

\subsection{Existing decoders}

\subsubsection{Random perturbation\label{subsec:Random-perturbation}}

A number of modified decoders have been presented in Ref. \citep{poulin2008iterative}
to address symmetric degeneracy errors. The best performing of these
is the random perturbation decoder, which attempts to break decoding
symmetries by randomizing the channel error probabilities. Initially,
decoding is attempted using a standard $\mathrm{GF}(4)$ decoder.
If this results in $\hat{\boldsymbol{z}}=\boldsymbol{z}$, then decoding
is complete. Otherwise, if $\hat{\boldsymbol{z}}\neq\boldsymbol{z}$,
then decoding is iteratively reattempted with modified error probabilities
until either decoding is successful or a maximum number of attempts
$N$ is reached. In each decoding attempt a frustrated check is selected.
This is a check node $c_{i}$ such that \textbf{$\hat{z}_{i}\neq z_{i}$}.
The channel probabilities of all qubits $j\in\mathcal{M}(i)$ involved
in this check are then perturbed (up to normalization) as follows:
\begin{align}
P(E_{j} & =I)\rightarrow P(E_{j}=I),\\
P(E_{j} & =X)\rightarrow(1+\delta_{X})P(E_{j}=X),\\
P(E_{j} & =Y)\rightarrow(1+\delta_{Y})P(E_{j}=Y),\\
P(E_{j} & =Z)\rightarrow(1+\delta_{Z})P(E_{j}=Z).
\end{align}
Here $\delta_{X}$, $\delta_{Y}$, and $\delta_{Z}$ are realizations
of a random variable that is uniformly distributed over $[0,\delta]$,
where $\delta$ is called the perturbation strength. The increasing
of non-identity error probabilities is motivated by the empirical
observation that the decoder is naturally too biased towards the trivial
error \citep{poulin2008iterative}.

\subsubsection{Enhanced feedback}

The enhanced feedback (EFB) decoder of Ref. \citep{wang2012enhanced},
which is specifically tailored for the depolarizing channel, behaves
somewhat similarly to the random perturbation decoder in that it also
iteratively reattempts decoding with modified channel probabilities.
Again, decoding is first attempted using a standard $\mathrm{GF}(4)$
decoder. If this results in $\hat{\boldsymbol{z}}=\boldsymbol{z}$,
then decoding is complete. If instead $\hat{\boldsymbol{z}}\neq\boldsymbol{z}$,
then a frustrated check $c_{i}$ is selected along with an involved
qubit $j\in\mathcal{M}(i)$. If $z_{i}=1$ but $\hat{z}_{i}=0$, then
the estimated error $\hat{E}$ commutes with the stabilizer generator
$M_{i}$ while the error $E$ anticommutes with $M_{i}$. To address
this, the channel probabilities for $E_{j}$ are adjusted such that
an anticommuting error is more likely than the commuting trivial error
that the decoder is naturally too biased towards. This adjustment
is
\begin{equation}
P(E_{j}=\sigma)\rightarrow\begin{cases}
\frac{p}{2} & \mathrm{if}\,\sigma=I,\,\mathrm{or}\,M_{i}^{(j)},\\
\frac{1-p}{2} & \mathrm{otherwise},
\end{cases}
\end{equation}
where $M_{i}^{(j)}$ is the $j$-th component of the generator $M_{i}$.
Conversely, if $z_{i}=0$ but $\hat{z}_{i}=1$, then the adjustment
is
\begin{equation}
P(E_{j}=\sigma)\rightarrow\begin{cases}
\frac{1-p}{2} & \mathrm{if}\,\sigma=I,\,\mathrm{or}\,M_{i}^{(j)},\\
\frac{p}{2} & \mathrm{otherwise}.
\end{cases}
\end{equation}
Decoding is then reattempted with these adjusted probabilities. If
this fails, then a different qubit $j\in\mathcal{M}(i)$ is selected
and the process is repeated. If all qubits involved in check $c_{i}$
have been exhausted and decoding is still unsuccessful, then a different
check is selected and the process continues. Again, decoding is halted
if a maximum number of attempts $N$ is reached.

\subsubsection{Supernodes}

The supernode decoder of Ref. \citep{babar2015fifteen} is a modification
of the $\mathrm{GF}(4)$ decoder for dual-containing CSS codes. Decoding
is performed on the factor graph corresponding to $\tilde{H}=\tilde{H}_{X}=\tilde{H}_{Z}$
with checks $c_{i}$ and $c_{i+m/2}$ grouped to form a single supernode.
The check node calculation is modified to
\begin{align}
\lambda_{i\rightarrow j}^{a}= & K\sum_{\boldsymbol{e}:e_{j}=a}\delta\left(\mathrm{tr}(\sum_{j'\in\mathcal{M}(i)}\bar{e_{j'}})=z_{Z}^{(i)}\right)\nonumber \\
 & \times\delta\left(\mathrm{tr}(\sum_{j'\in\mathcal{M}(i)}\omega\bar{e_{j'}})=z_{X}^{(i)}\right)\prod_{j'\in\mathcal{M}(i)\backslash j}\mu_{j'\rightarrow i}^{e_{j'}}.\label{eq:quantum check supernode pre}
\end{align}
Here $\boldsymbol{z}_{Z}$ contains the first $m/2$ values of $\boldsymbol{z}$
and $\boldsymbol{z}_{X}$ contains the last $m/2$ values; $z_{Z}^{(i)}$
and $z_{X}^{(i)}$ are the $i$-th values of $\boldsymbol{z}_{Z}$
and $\boldsymbol{z}_{X}$ respectively. Defining $\tilde{z}_{i}=\omega z_{Z}^{(i)}+z_{X}^{(i)}\in\mathrm{GF}(4)$,
the two constraints of Eq. (\ref{eq:quantum check supernode pre})
can be combined to give
\begin{equation}
\lambda_{i\rightarrow j}^{a}=K\sum_{\boldsymbol{e}:e_{j}=a}\delta\left(\sum_{j'\in\mathcal{M}(i)}e_{j'}=\tilde{z}_{i}\right)\prod_{j'\in\mathcal{M}(i)\backslash j}\mu_{j'\rightarrow i}^{e_{j'}}.\label{eq:quantum check supernode-1}
\end{equation}
Note that this is of the same form as the classical check to error
message given in Eq. (\ref{eq:check to error message}), and it can
therefore be computed using the same Fourier transform approach. The
effect of combining nodes into supernodes is twofold. Firstly, it
reduces decoding complexity by halving the number of check node calculations.
Secondly, it can improve decoder performance as it reduces the number
of $4$-cycles present in the factor graph. Note that random perturbation
and EFB can also be implemented using an underlying supernode decoder
rather than a standard $\mathrm{GF}(4)$ decoder.

\subsection{New decoders}

\subsubsection{Adjusted\label{subsec:Binary-adjusted}}

The first decoder we propose is the adjusted decoder for CSS codes.
This is a $\mathrm{GF}(2)$ based decoder that aims to reintroduce
the correlations between $X$ and $Z$ errors that are lost when using
a standard $\mathrm{GF}(2)$ decoder. Initially, decoding is attempted
using a standard $\mathrm{GF}(2)$ decoder. If this is successful,
then decoding is complete. If both $H_{Z}\hat{\boldsymbol{e}}_{X}=\hat{\boldsymbol{z}}_{X}\neq\boldsymbol{z}_{X}$
and $H_{X}\hat{\boldsymbol{e}}_{Z}=\hat{\boldsymbol{z}}_{Z}\neq\boldsymbol{z}_{Z}$,
then the adjusted decoder also halts. However, if one of $\hat{\boldsymbol{z}}_{X}=\boldsymbol{z}_{X}$
or $\hat{\boldsymbol{z}}_{Z}=\boldsymbol{z}_{Z}$, then we reattempt
decoding for the incorrect component using channel probabilities that
are adjusted according to Eqs. (\ref{eq:cond prob 1}) to (\ref{eq:cond prob 4}).
In particular, if $\hat{\boldsymbol{z}}_{X}=\boldsymbol{z}_{X}$ but
$\hat{\boldsymbol{z}}_{Z}\neq\boldsymbol{z}_{Z}$, then the adjustment
is
\begin{equation}
P(e_{Z}^{(j)}=1)\rightarrow\begin{cases}
\frac{p_{Y}}{p_{X}+p_{Y}} & \text{if}\,\hat{e}_{X}^{(j)}=1,\\
\frac{p_{Z}}{1-(p_{X}+p_{Y})} & \text{if}\,\hat{e}_{X}^{(j)}=0.
\end{cases}\label{eq:adjusted Z probabilities}
\end{equation}
Alternatively, if $\hat{\boldsymbol{z}}_{Z}=\boldsymbol{z}_{Z}$ but
$\hat{\boldsymbol{z}}_{X}\neq\boldsymbol{z}_{X}$, then the adjustment
is 
\begin{equation}
P(e_{X}^{(j)}=1)\rightarrow\begin{cases}
\frac{p_{Y}}{p_{Y}+p_{Z}} & \text{if}\,\hat{e}_{Z}^{(j)}=1,\\
\frac{p_{X}}{1-(p_{Y}+p_{Z})} & \text{if}\,\hat{e}_{Z}^{(j)}=0.
\end{cases}
\end{equation}
We note that the adjusted decoder presented here is similar to the
decoder presented for the depolarizing channel in Ref. \citep{delfosse2014decoding}.
The decoder of Ref. \citep{delfosse2014decoding} first attempts decoding
of the $X$ component using standard channel probabilities. If this
is successful, then decoding is attempted for the $Z$ components
using the modified probabilities of Eq. (\ref{eq:adjusted Z probabilities}).

\subsubsection{Augmented}

The second decoder we propose is the augmented decoder, which was
first presented in Ref. \citep{rigby2018augmented} for classical
binary codes. An augmented decoder for QLDPC codes can be based on
a $\mathrm{GF}(4)$ decoder for any code, a $\mathrm{GF}(2)$ decoder
for a CSS code, or a supernode decoder for a dual-containing CSS code.
The simplest of these cases is when the underlying decoder is a $\mathrm{GF}(4)$
decoder. In this case, decoding is initially attempted using a standard
$\mathrm{GF}(4)$ decoder with a standard $\mathrm{GF}(4)$ parity-check
matrix $H$. If this is unsuccessful, then decoding is reattempted
using a randomly generated augmented parity-check matrix
\begin{equation}
H_{A}=\left(\begin{array}{c}
H\\
H_{\delta}
\end{array}\right).
\end{equation}
$H_{\delta}$ is comprised of a subset of rows selected at random
from $H$. The fraction of rows selected is dictated by the augmentation
density $\delta$. The syndrome used for decoding is
\begin{equation}
\boldsymbol{z}_{A}=\left(\begin{array}{c}
\boldsymbol{z}\\
\boldsymbol{z}_{\delta}
\end{array}\right),
\end{equation}
where $\boldsymbol{z}$ is the measured syndrome and $\boldsymbol{z}_{\delta}$
contains the syndrome values corresponding to the rows selected to
form $H_{\delta}$. Decoding is iteratively reattempted using different
augmented matrices until either decoding is successful or a maximum
number of attempts $N$ is reached. Note that duplicating rows results
in a duplication of the corresponding check nodes in the factor graph.

The behavior of a supernode based augmented decoder is very similar.
In this case the augmented parity-check matrices are of the form
\begin{equation}
H_{A}=\left(\begin{array}{c}
\tilde{H}\\
\tilde{H}_{\delta}
\end{array}\right),\label{eq:augmented supernode pcm}
\end{equation}
where $\tilde{H}_{\delta}$ consists of the rows selected from $\tilde{H}=\tilde{H}_{X}=\tilde{H}_{Z}$.
The augmented syndrome is
\begin{equation}
\boldsymbol{z}_{A}=\left(\begin{array}{c}
\boldsymbol{z}_{Z}\\
\boldsymbol{z}_{Z\delta}\\
\boldsymbol{z}_{X}\\
\boldsymbol{z}_{X\delta}
\end{array}\right),
\end{equation}
where the values of $\boldsymbol{z}_{Z\delta}$ and $\boldsymbol{z}_{X\delta}$
are taken from $\boldsymbol{z}_{Z}$ and $\boldsymbol{z}_{X}$ respectively
according to the rows selected for repetition.

In the $\mathrm{GF}(2)$ case two augmented decoders are used, one
for the $X$ component and one for the $Z$ component. The augmented
parity-check matrices used by the $X$ decoder are of the form 
\begin{equation}
H_{A}=\left(\begin{array}{c}
\tilde{H}_{Z}\\
\tilde{H}_{Z\delta}
\end{array}\right),\label{eq:augmented supernode pcm-1}
\end{equation}
and the augmented syndrome is
\begin{equation}
\boldsymbol{z}_{A}=\left(\begin{array}{c}
\boldsymbol{z}_{X}\\
\boldsymbol{z}_{X\delta}
\end{array}\right).
\end{equation}
The syndrome and augmented parity-check matrices used by the $Z$
decoder are of the same form.

In all three cases ($\mathrm{GF}(2)$, $\mathrm{GF}(4)$, and supernode),
decoding with an augmented parity-check matrix $H_{A}$ is equivalent
to running a slightly altered belief propagation algorithm using the
standard parity-check matrix $H$. We define the function $r$ such
that
\begin{equation}
r(i)=\begin{cases}
1 & \mathrm{if}\,c_{i}\,\mathrm{duplicated\,in}\,H_{A},\\
0 & \mathrm{otherwise}.
\end{cases}
\end{equation}
 Decoding with $H_{A}$ is then equivalent to decoding using $H$
with the marginal marginal probability approximation of Eq. (\ref{eq:marignal probability estimate})
changed to
\begin{equation}
\hat{P}(e_{j}=a|\boldsymbol{z})=KP(e_{j}=a)\prod_{i\in\mathcal{N}(j)}(\lambda_{i\rightarrow j}^{a})^{1+r(i)},\label{eq:marignal probability estimate-1}
\end{equation}
and the error to check message of Eq. (\ref{eq:error to check message})
changed to
\begin{equation}
\mu_{j\rightarrow i}^{a}=KP(e_{j}=a)(\lambda_{i\rightarrow j}^{a})^{r(i)}\prod_{i'\in\mathcal{N}(j)\backslash i}(\lambda_{i'\rightarrow j}^{a})^{1+r(i')}.\label{eq:error to check message-1}
\end{equation}
As a result of this equivalence, we can consider one iteration of
an augmented decoder to be of the same complexity as one iteration
of the underlying decoder. This formulation also gives some insight
into the effect of decoding with an augmented parity-check matrix.
It can be seen that repeating a check has the effect of increasing
its influence in estimating the error. Furthermore, the message $\mu_{j\rightarrow i}$
is now no longer independent of the message $\lambda_{i\rightarrow j}$
if $c_{i}$ is duplicated. This amplification and feedback will alter
the convergence of the marginal probability estimates. This altered
convergence can help the decoder to give a different (and hopefully
correct) error estimate.

\subsubsection{Combined}

The third decoder we propose combines the augmented $\mathrm{GF}(2)$
and adjusted decoders for CSS codes. Initially, standard $\mathrm{GF}(2)$
decoding is attempted. If this is successful, then decoding is complete.
If both $\hat{\boldsymbol{z}}_{X}\neq\boldsymbol{z}_{X}$ and $\hat{\boldsymbol{z}}_{Z}\neq\boldsymbol{z}_{Z}$,
then we reattempt decoding for the $X$ component using augmented
parity-check matrices up to $N$ times. If this is unsuccessful, then
we repeat this procedure for the $Z$ component. If we still have
$\hat{\boldsymbol{z}}_{X}\neq\boldsymbol{z}_{X}$ and $\hat{\boldsymbol{z}}_{Z}\neq\boldsymbol{z}_{Z}$,
then decoding halts. However, if one of $\hat{\boldsymbol{z}}_{X}=\boldsymbol{z}_{X}$
or $\hat{\boldsymbol{z}}_{Z}=\boldsymbol{z}_{Z}$ (either from the
initial decoding or after attempting decoding with augmented parity-check
matrices if required), then we reattempt decoding for the unsatisfied
component with adjusted channel error probabilities as outlined in
Sec. \ref{subsec:Binary-adjusted}. If this is unsuccessful, then
decoding for this component will be reattempted with augmented parity-check
matrices up to $N$ times using the same adjusted probabilities.

\section{Simulation results\label{sec:Simulation-results}}

\subsection{Bicycle\label{subsec:Bicycle}}

The first code we have considered is a $[[400,200]]$ bicycle code
of Ref. \citep{mackay2004sparse}. Bicycle codes are dual-containing
CSS codes that are constructed by first generating an $n/2\times n/2$
binary circulant matrix $A$ with row weight $w/2$. $A$ is used
to define the $n/2\times n$ matrix $H_{0}=[\begin{array}{cc}
A & A^{T}\end{array}]$ from which $(n-m)/2$ rows are removed to give $\tilde{H}$ (following
the heuristic that column weight should be kept as uniform as possible).
Taking $\tilde{H}_{X}=\tilde{H}_{Z}=\tilde{H}$ defines the $\mathrm{GF}(2)$
and $\mathrm{GF}(4)$ parity-check matrices according to Eqs. (\ref{eq:css pcm})
and (\ref{eq:gf4 pcm}) respectively. The associated stabilizer code
will have $k\geq n-m$, with equality when the parity-check matrix
is full rank (this is the case for our code). Removing rows from $H_{0}$
corresponds to removing stabilizer generators of weight $w$. Unless
a removed row belongs to the span of the remaining rows, which is
unlikely, the removed generator will be in $N(\tilde{\mathcal{S}})\backslash\mathcal{\tilde{S}}$.
A bicycle code's distance is therefore upper bounded by $w$ (we have
chosen $w=20$ for our code).

\subsubsection{Depolarizing channel}

\begin{figure}
\includegraphics[scale=0.55]{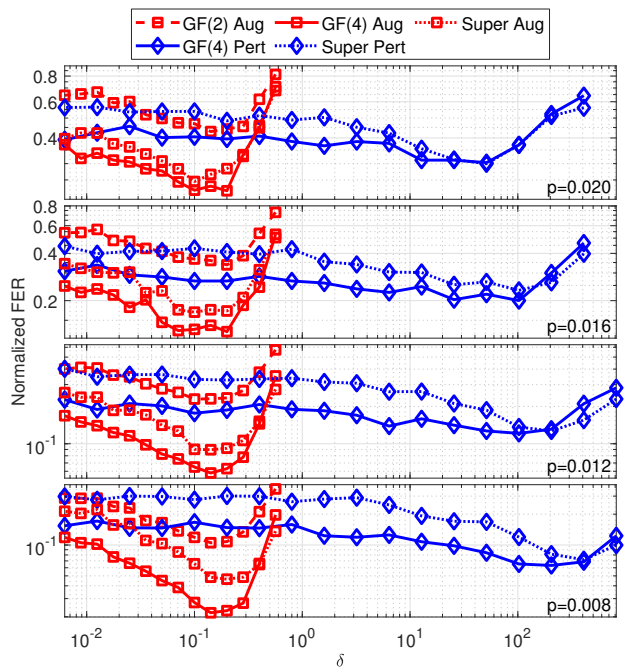}\caption{\label{n400 dens fig}The effect of augmentation density and random
perturbation strength on decoder performance for the $[[400,200]]$
bicycle code on the depolarizing channel. Each decoder uses $N=10$
maximum attempts.}
\end{figure}

We first consider the depolarizing channel. Both the augmented and
random perturbation decoders have a tunable parameter $\delta$, which
controls the augmentation density and perturbation strength respectively.
As shown for classical codes in Ref. \citep{rigby2018augmented},
this $\delta$ value can have a significant impact on the performance
of an augmented decoder. We observe the same behavior for both augmented
and random perturbation decoders in the quantum case as shown in Fig.
\ref{n400 dens fig}. Here decoders with $N=10$ maximum decoding
attempts and varying $\delta$ have been tested at four different
depolarizing probabilities (we use a maximum of $I_{\max}=100$ iterations
per attempt for every decoder in this paper). The vertical axis gives
normalized FER, which is the modified decoder's FER divided by the
underlying (standard) decoder's FER. Note that each data point in
these figures, as well as all other figures presented in this paper,
corresponds to at least $100$ decoding errors. Based on these results,
we have selected values of $\delta=0.1$ for the augmented $\mathrm{GF}(2)$
decoder, $\delta=0.15$ for the augmented $\mathrm{GF}(4)$ and supernode
decoders, $\delta=100$ for the random perturbation $\mathrm{GF}(4)$
decoder, and $\delta=200$ for the random perturbation supernode decoder.
Note that the $\delta$ value we use for the combined decoder is always
the same as the value used for the augmented $\mathrm{GF}(2)$ decoder.

\begin{figure}
\includegraphics[scale=0.55]{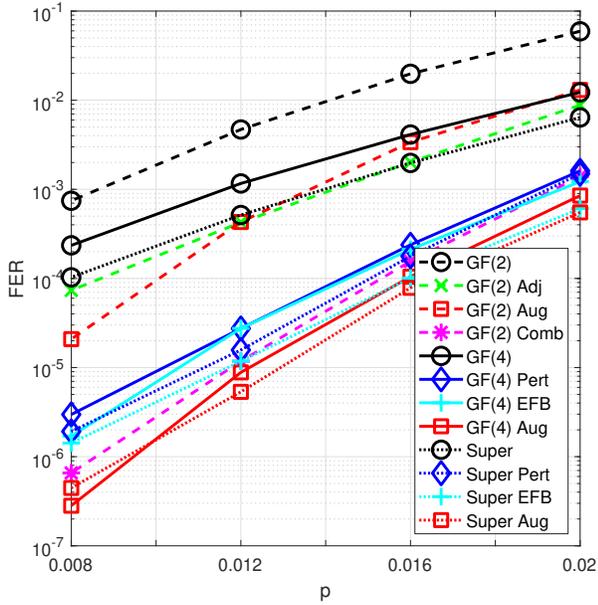}\caption{\label{n400 FER fig}FER performance of decoders with $N=100$ attempts
(where applicable) for the $[[400,200]]$ bicycle code on the depolarizing
channel.}
\end{figure}

\begin{figure}
\includegraphics[scale=0.55]{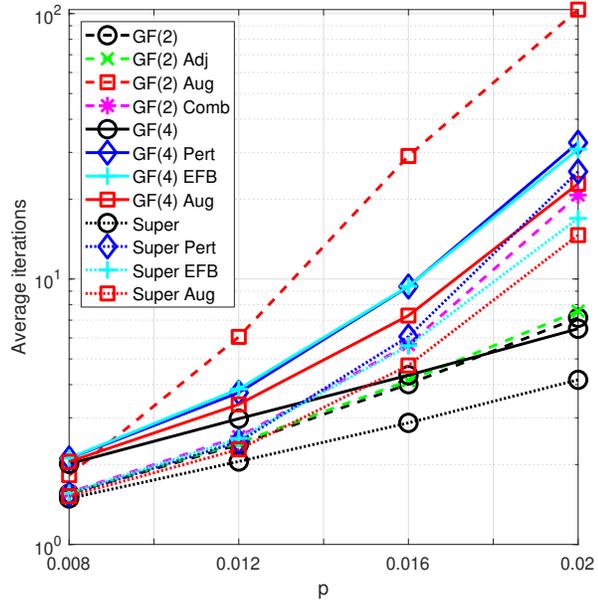}\caption{\label{n400 iters fig}Average number of iterations required by decoders
with $N=100$ attempts (where applicable) for the $[[400,200]]$ bicycle
code on the depolarizing channel.}
\end{figure}

We have tested all of the decoders outlined in Sec. \ref{sec:Modified-decoders}
on this code. The random perturbation, EFB, augmented, and combined
decoders all use $N=100$ attempts. The FER performance of these decoders
is shown in Fig. \ref{n400 FER fig}, and the average number of iterations
required by each of them is shown in Fig. \ref{n400 iters fig}. It
can be seen that the standard supernode decoder outperforms the standard
$\mathrm{GF}(4)$ decoder, which in turn outperforms the standard
$\mathrm{GF}(2)$ decoder. Furthermore, the supernode decoder requires
fewer iterations on average than the standard $\mathrm{GF}(4)$ or
$\mathrm{GF}(2)$ decoders (the number of iterations used by a $\mathrm{GF}(2)$
based decoder is taken to be the number used by one of the two constituent
decoders). However, note that comparing the number of iterations used
by these different decoders, or indeed modified decoders based on
different underlying decoders, is not particularly meaningful as their
iterations are of differing complexity. The adjusted decoder can be
seen to give a FER similar to the standard supernode decoder at the
cost of a negligible increase in required iterations compared to the
standard $\mathrm{GF}(2)$ decoder. This FER performance suggests
that the adjusted decoder is successful in reintroducing the correlation
between the $X$ and $Z$ error components. The random perturbation
and EFB decoders based on either $\mathrm{GF}(4)$ or supernode decoders
have similar FER performance and require a near-identical number of
iterations on average. The augmented $\mathrm{GF}(4)$ and supernode
decoders outperform both the random perturbation and EFB decoders
while requiring a lower number of iterations on average. The augmented
$\mathrm{GF}(2)$ decoder does give a reasonable FER reduction compared
to the standard $\mathrm{GF}(2)$ decoder, but it is outperformed
by all modified $\mathrm{GF}(4)$ and supernode decoders. However,
the combined decoder gives a FER lower than the random perturbation
and EFB decoders. Furthermore, it also requires fewer iterations on
average than the augmented $\mathrm{GF}(2)$ decoder. All of the decoding
errors we have observed for this code are detected errors; that is,
they are due to a failing of the decoder rather than the code's distance.

\begin{figure}
\includegraphics[scale=0.55]{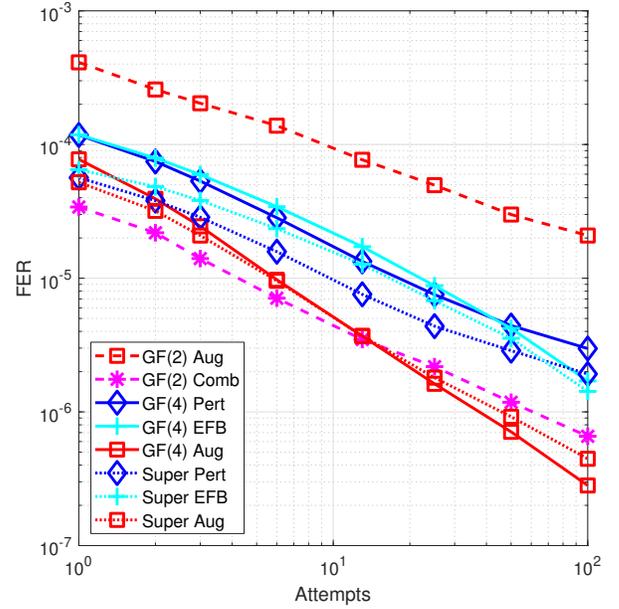}\caption{\label{n400 attempts fig}FER performance of decoders at $p=0.008$
with a varying number of decoding attempts for the $[[400,200]]$
bicycle code on the depolarizing channel.}
\end{figure}

Fig. \ref{n400 attempts fig} shows the effect of the maximum number
of decoding attempts on the performance of the augmented, combined,
random perturbation, and EFB decoders at a depolarizing probability
of $p=0.008$. For all decoders, the FER reduction with an increasing
maximum number of attempts is approximately linear on a log-log plot.
This suggests that we could continue to reduce the FER by increasing
the maximum number of attempts beyond $N=100$. It can be seen that
the augmented and combined decoders only require approximately $N=25$
maximum attempts to match the performance of random perturbation and
EFB decoders with $N=100$.

\subsubsection{$XZ$ channel}

To isolate the effect of augmentation in the $\mathrm{GF}(2)$ case,
we have repeated the analysis of the previous section for the $XZ$
channel. As previously noted, the $X$ and $Z$ error components occur
independently for this channel, therefore, there are no correlations
to be ignored when using a $\mathrm{GF}(2)$ based decoder. As a result,
the adjusted and combined decoders will give no performance increase
over the standard $\mathrm{GF(2)}$ and augmented $\mathrm{GF}(2)$
decoders respectively. While we have still employed the random perturbation
decoder for comparison on this channel, we have not used the EFB decoder
as it is specifically tailored to the depolarizing channel.

\begin{figure}
\includegraphics[scale=0.55]{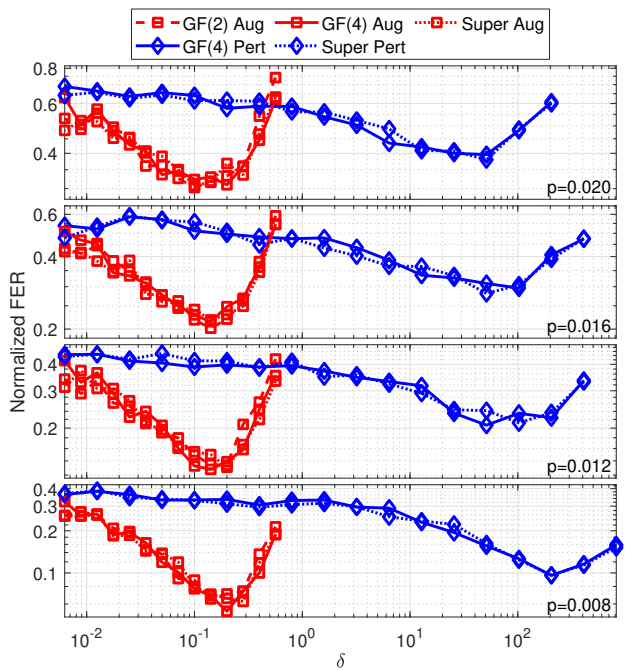}\caption{\label{n400 dens fig XZ}The effect of augmentation density and random
perturbation strength on decoder performance for the $[[400,200]]$
bicycle code on the $XZ$ channel. Each decoder uses $N=10$ maximum
attempts.}
\end{figure}

Again, we first tune the augmentation density and and random perturbation
strength using decoders with $N=10$ as shown in Fig. \ref{n400 dens fig XZ}.
It can be seen that the optimal value of $\delta$ is essentially
independent of the underlying decoder. As such, we have selected a
value of $\delta=0.15$ for all augmented decoders and $\delta=100$
for both of the random perturbation decoders. Note that these are
the same values we have used for the $\mathrm{GF(4)}$ based decoders
in the depolarizing case.

\begin{figure}
\includegraphics[scale=0.55]{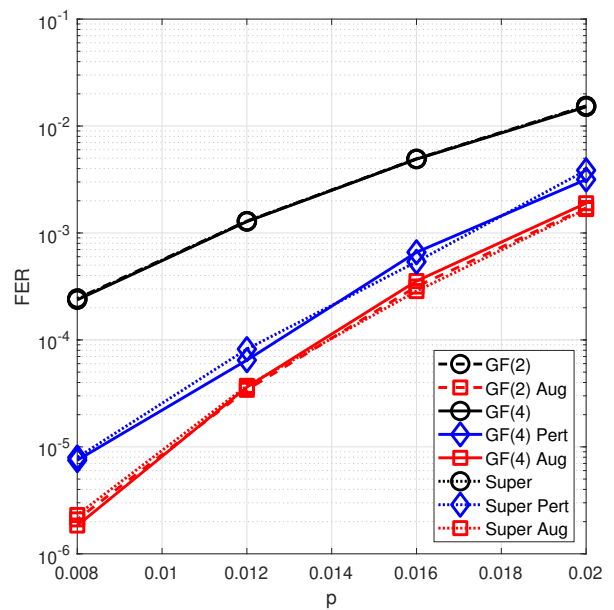}\caption{\label{n400 FER fig XZ}FER performance of decoders with $N=100$
attempts (where applicable) for the $[[400,200]]$ bicycle code on
the $XZ$ channel.}
\end{figure}

\begin{figure}
\includegraphics[scale=0.55]{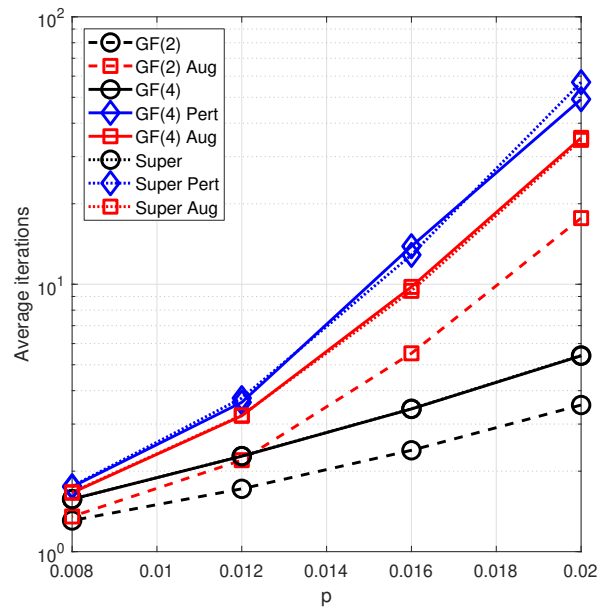}\caption{\label{n400 iters fig XZ}Average number of iterations required by
decoders with $N=100$ attempts (where applicable) for the $[[400,200]]$
bicycle code on the $XZ$ channel.}
\end{figure}

The FER performance and average required iterations for decoders with
$N=100$ maximum attempts are shown in Figs. \ref{n400 FER fig XZ}
and \ref{n400 iters fig XZ} respectively. It can be seen that the
standard $\mathrm{GF}(2)$, $\mathrm{GF}(4)$, and supernode decoders
all exhibit near-identical performance on the $XZ$ channel. That
the $\mathrm{GF}(2)$ and supernode decoders yield the same FER is
unsurprising and is consistent with the similar performance of the
adjusted and supernode decoders on the depolarizing channel. The performance
of the $\mathrm{GF}(4)$ decoder suggests that the $4$-cycles involving
one row from $\tilde{H}_{X}$ and one row from $\tilde{H}_{Z}$ have
no effect on decoding performance when the error components are independent.
The performance of the augmented and random perturbation decoders
is also largely independent of the underlying decoder. Furthermore,
the relative performance of the decoders is very similar to that observed
for the $\mathrm{GF}(4)$ based decoders in the depolarizing case,
with the augmented decoders outperforming the random perturbation
decoders.

\begin{figure}
\includegraphics[scale=0.55]{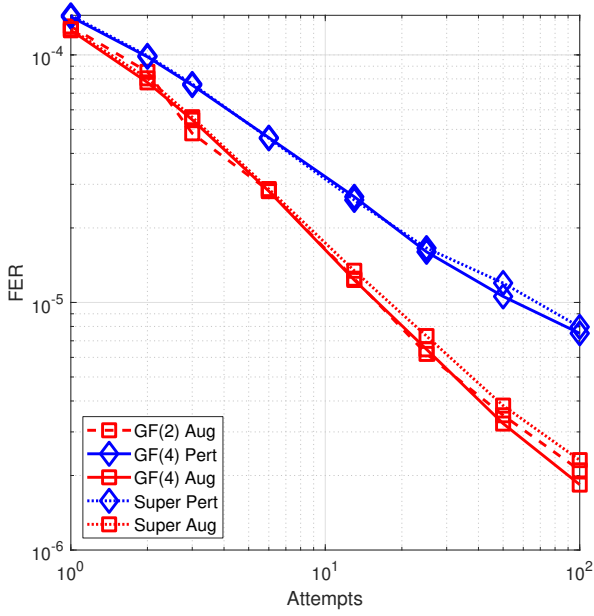}\caption{\label{n400 attempts fig XZ}FER performance of decoders at $p=0.008$
with a varying number of decoding attempts for the $[[400,200]]$
bicycle code on the $XZ$ channel.}
\end{figure}

The effect of the maximum number of decoding attempts on decoder performance
is shown for $p=0.008$ in Fig. \ref{n400 attempts fig XZ}. Unsurprisingly,
the performance of the augmented and random perturbation decoders
remains largely independent of the underlying decoder over the range
of $N$ values tested. Furthermore, the relative performance is very
similar to that exhibited by the $\mathrm{GF}(4)$ based decoders
in the depolarizing case, with the augmented decoders only requiring
approximately $N=25$ maximum attempts to match the performance of
the random perturbation decoders with $N=100$.

We have tested the performance of decoders on the $XZ$ channel for
all four CSS codes considered in this paper. However, we omit the
results for the other three codes as they all follow the same trend
outlined here. That is, the performance of decoders is essentially
independent of the underlying decoder, and the relative performance
of the augmented and random perturbation decoders is very similar
to that exhibited by the $\mathrm{GF}(4)$ based decoders in the depolarizing
case. 

\subsection{BIBD\label{subsec:BIBD}}

The second code we have considered is a $[[610,490]]$ balanced incomplete
block design (BIBD) code from Ref. \citep{djordjevic2008quantum}.
Like the bicycle code, this is also a dual-containing CSS code. A
BIBD $(X,\mathcal{B})$, where $X=\{x_{1},\dots,x_{v}\}$ and $\mathcal{B}=\{B_{1},\dots,B_{b}\}$,
is a collection of $b$ subsets (blocks) of size $k$ that are drawn
from a set $X$ containing $v$ elements. Each pair of elements occurs
in $\lambda$ of the blocks, and every element occurs in $r$ blocks.
The $v\times b$ $\mathrm{GF}(2)$ incidence matrix $A$ of $(X,\mathcal{B})$
has elements
\begin{equation}
A_{ij}=\begin{cases}
1 & x_{i}\in B_{j},\\
0 & x_{i}\notin B_{j}.
\end{cases}
\end{equation}
If $\lambda$ is even, then $A$ will satisfy $AA^{T}=0$ as any two
rows will overlap an even number of times. As such, taking $\tilde{H}=\tilde{H}_{X}=\tilde{H}_{Z}=A$
defines a dual-containing CSS code. The BIBD that we have selected
follows the construction of Ref. \citep{bose1939construction}. If
$6t+1$ is a prime or prime power and $\alpha$ is a primitive element
of $\mathrm{GF}(6t+1)$, then a BIBD $(\mathrm{GF}(6t+1),\mathcal{B})$
can be constructed with $v=6t+1$, $b=t(6t+1)$, $r=4t$, $k=4$,
and $\lambda=2$. To do this, $t$ base blocks $\tilde{B}_{i}$ are
constructed for $0\leq i\leq t-1$ with
\begin{equation}
\tilde{B}_{i}=\{0,\alpha^{i},\alpha^{2t+i},\alpha^{4t+i}\}.
\end{equation}
$6t+1$ blocks of the form $\tilde{B}_{i}+\beta=\{\beta,\alpha^{i}+\beta,\alpha^{2t+i}+\beta,\alpha^{4t+i}+\beta\}$,
where $\beta\in\mathrm{GF}(6t+1)$, can then be constructed from each
base block. This gives a total of $t(6t+1)$ blocks and a corresponding
incidence matrix of the form
\begin{equation}
\tilde{H}=A=\left(\begin{array}{cccc}
A_{1} & A_{2} & \cdots & A_{t}\end{array}\right).\label{eq:bibd adj}
\end{equation}
Here each $A_{i}$ is a $(6t+1)\times(6t+1)$ circulant matrix of
weight $k=4$. We have selected $t=10$ and $\alpha=2$ for our code.

\begin{figure}
\includegraphics[scale=0.55]{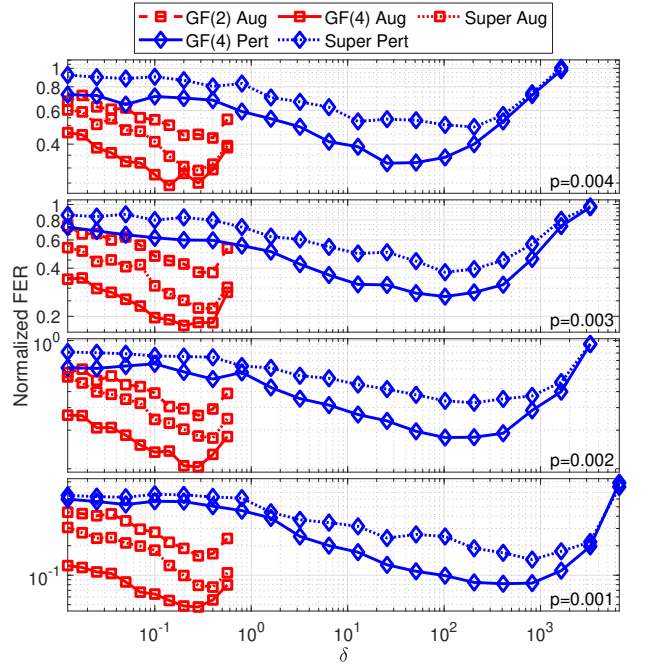}\caption{\label{n610 BIBD dens fig}The effect of augmentation density and
random perturbation strength on decoder performance for the $[[610,490]]$
BIBD code on the depolarizing channel. Each decoder uses $N=10$ maximum
attempts.}
\end{figure}

The results presented for this code and all codes that follow are
on the depolarizing channel. The effect of augmentation density and
random perturbation strength for decoders with $N=10$ on this code
is shown in Fig. \ref{n610 BIBD dens fig}. Based on these results,
we have selected values of $\delta=0.3$ for all augmented decoders,
$\delta=200$ for the random perturbation $\mathrm{GF}(4)$ decoder,
and $\delta=400$ for the random perturbation supernode decoder.

\begin{figure}
\includegraphics[scale=0.55]{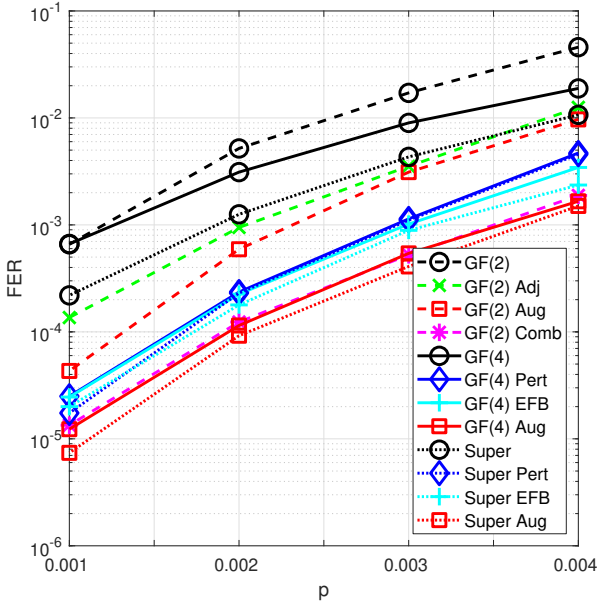}\caption{\label{n610 BIBD FER fig}FER performance of decoders with $N=100$
attempts (where applicable) for the $[[610,490]]$ BIBD code on the
depolarizing channel.}
\end{figure}

\begin{figure}
\includegraphics[scale=0.55]{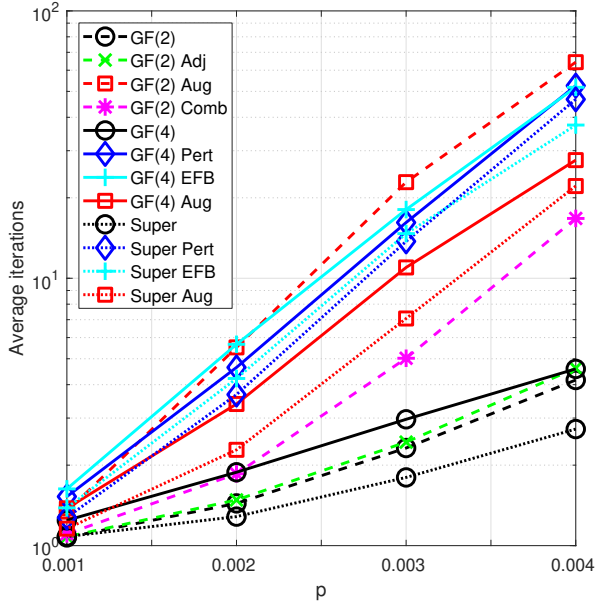}\caption{\label{n610 BIBD iters fig}Average number of iterations required
by decoders with $N=100$ attempts (where applicable) for the $[[610,490]]$
BIBD code on the depolarizing channel.}
\end{figure}

The FER performance and average required iterations for decoders with
$N=100$ maximum attempts are shown in Figs. \ref{n610 BIBD FER fig}
and \ref{n610 BIBD iters fig} respectively. The results here are
quite similar to those for the bicycle code. Again, the adjusted decoder
gives performance similar to that of the supernode decoder. Furthermore,
the random perturbation and EFB decoders perform similarly to one
another. The augmented $\mathrm{GF}(2)$ decoder is outperformed by
all modified $\mathrm{GF}(4)$ and supernode decoders. The combined,
augmented $\mathrm{GF}(4)$, and augmented supernode decoders again
outperform the random perturbation and EFB decoders. Overall there
is less spread in the performance of the decoders on this BIBD code.
This can be attributed to the fact that a large fraction of decoding
errors are undetected. For example, approximately $65\%$ of the errors
exhibited by the augmented supernode decoder at $p=0.001$ are undetected.
This abundance of undetected errors suggests that decoding is being
limited by the code's distance $d\leq5$ (this value is based on the
lowest weight element of $N(\tilde{\mathcal{S}})\backslash\mathcal{\tilde{S}}$
that we have observed).

\begin{figure}
\includegraphics[scale=0.55]{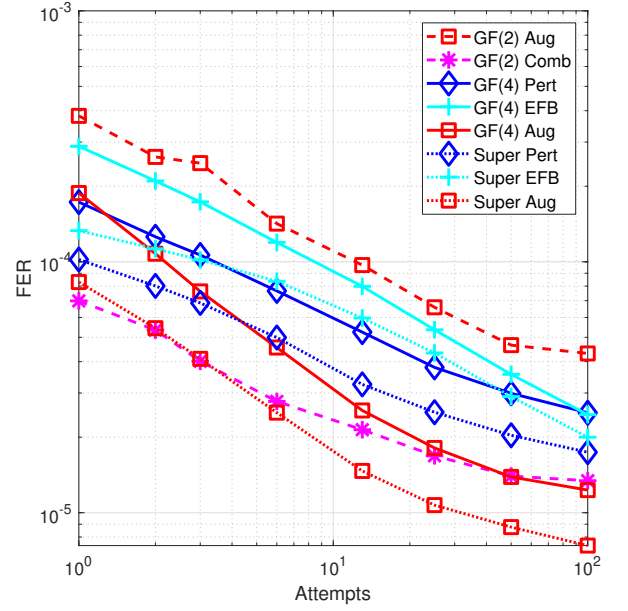}\caption{\label{n610 BIBD attempts fig}FER performance of decoders at $p=0.001$
with a varying number of decoding attempts for the $[[610,490]]$
BIBD code on the depolarizing channel.}
\end{figure}

The effect of these undetected errors can also be seen in Fig. \ref{n610 BIBD attempts fig}.
For the bicycle code, the reduction in FER with increasing maximum
number of iterations was approximately linear on a log-log plot. However,
the reduction in FER for the BIBD code can be seen to taper off; that
is, increasing the maximum number of attempts has diminishing returns.
Partially as a result of this, we only require approximately $N=10$
maximum attempts for our augmented supernode decoder to match the
performance of the random perturbation and EFB decoders with $N=100$.

\subsection{Quasi-cyclic\label{subsec:Quasi-cyclic}}

The third code we have considered is a $[[506,240]]$ quasi-cyclic
code from Ref. \citep{hagiwara2007quantum}. Unlike the first two
codes, this is a non-dual-containing CSS code. The parity-check submatrices
$\tilde{H}_{X}$ and $\tilde{H}_{Z}$ can be defined in terms of base
matrices $\mathcal{H}_{X}$ and $\mathcal{H}_{Z}$ respectively whose
elements belong to the set $\{0,1,\dots,P-1\}$. $\tilde{H}_{X}$
($\tilde{H}_{Z}$) is then constructed by replacing each element of
$\mathcal{H}_{X}$ ($\mathcal{H}_{Z}$) with a $P\times P$ identity
matrix shifted circularly to the right by an amount given by the replaced
element. The base matrix construction of Ref. \citep{hagiwara2007quantum}
gives a parity-check matrix that satisfies $\tilde{H}_{Z}\tilde{H}_{X}^{T}=0$;
it also ensures that the factor graphs associated with $\tilde{H}_{X}$
and $\tilde{H}_{Z}$ are free of $4$-cycles. These base matrices
are constructed from a so-called ``perfume'' (perfect fulfillment).
Let $\mathbb{Z}_{P}$ be the set of integers $\{0,1,\dots,P-1\}$
with addition, subtraction, and multiplication modulo $P$. $\mathbb{Z}_{P}^{*}$
is then the abelian multiplicative group $\mathbb{Z}_{P}^{*}=\{z\in\mathbb{Z}_{P}:\gcd(z,P)=1\}$.
For positive integers $P$ and $\sigma$, $\sigma$ is a fulfillment
of $P$ if $\sigma$ is coprime to $P$ and $1-\sigma^{i}$ is coprime
to $P$ for $1\leq i<\mathrm{ord}(\sigma)$. Here $\mathrm{ord}(\sigma)$
is the order of $\sigma$ in $\mathbb{Z}_{P}^{*}$. A triple of positive
integers $(P,\sigma,\tau)$ is a perfume if $\sigma$ is a fulfillment
of $P$, $\tau$ is coprime to $P$, and $\tau\notin\{\sigma,\sigma^{2},\dots,\sigma^{\mathrm{ord}(\sigma)}\}$.
Letting $L=2\mathrm{ord}(\sigma)$, we define

\begin{equation}
c_{jl}=\begin{cases}
\sigma^{-j+l} & \mathrm{if}\,0\leq l<\frac{L}{2},\\
\tau\sigma^{-j+l} & \mathrm{if}\,\frac{L}{2}\leq l\leq L,
\end{cases}
\end{equation}
and

\begin{equation}
d_{kl}=\begin{cases}
-\tau\sigma^{k-l} & \mathrm{if}\,0\leq l<\frac{L}{2},\\
-\sigma^{k-l} & \mathrm{if}\,\frac{L}{2}\leq l\leq L.
\end{cases}
\end{equation}
Indexing from zero, these are the elements of the $J\times L$ and
$K\times L$ base matrices $\tilde{H}_{X}$ and $\tilde{H}_{Z}$ respectively
where $1\leq J,K\leq L/2$. To construct our code, we have used the
perfume $(23,8,20)$ (this gives $L=22$) and have chosen $J=K=6$.

\begin{figure}
\includegraphics[scale=0.55]{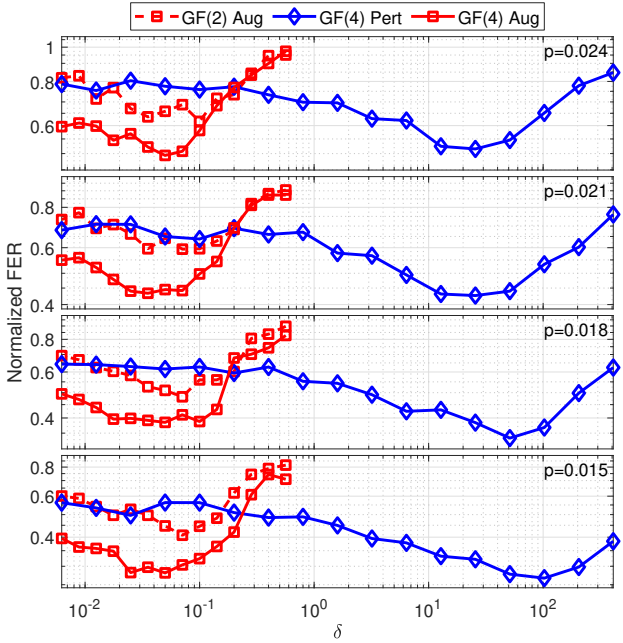}\caption{\label{n506 QC dens fig}The effect of augmentation density and random
perturbation strength on decoder performance for the $[[506,240]]$
quasi-cyclic code on the depolarizing channel. Each decoder uses $N=10$
maximum attempts.}
\end{figure}

The effect of augmentation density and random perturbation strength
for decoders with $N=10$ on this code is shown in Fig. \ref{n506 QC dens fig}.
Note that for this code we can only use $\mathrm{GF}(2)$ and $\mathrm{GF}(4)$
based decoders as it is not dual-containing. Based on these results,
we have selected values of $\delta=0.07$ for the augmented $\mathrm{GF}(2)$
decoder, $\delta=0.05$ for the augmented $\mathrm{GF}(4)$ decoder,
and $\delta=50$ for the random perturbation $\mathrm{GF}(4)$ decoder.

\begin{figure}
\includegraphics[scale=0.55]{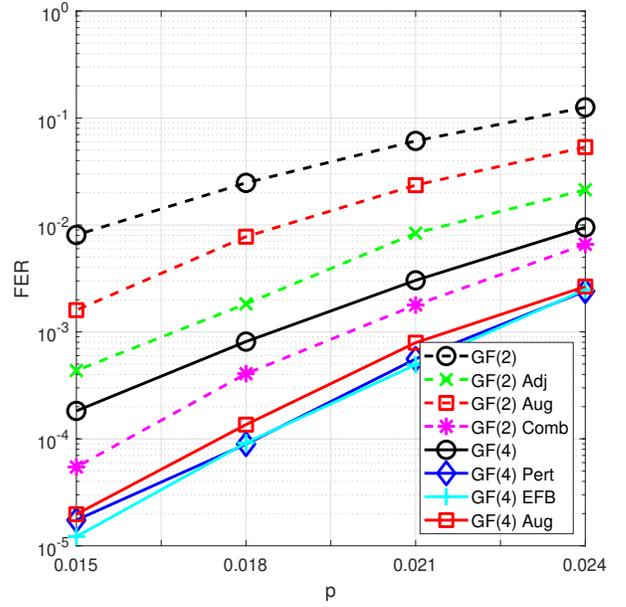}\caption{\label{n506 QC FER fig}FER performance of decoders with $N=100$
attempts (where applicable) for the $[[506,240]]$ quasi-cyclic code
on the depolarizing channel.}
\end{figure}

\begin{figure}
\includegraphics[scale=0.55]{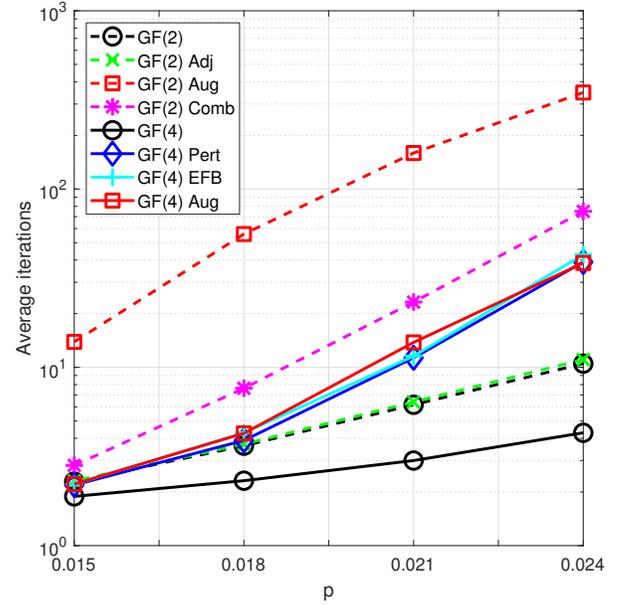}\caption{\label{n506 QC iters fig}Average number of iterations required by
decoders with $N=100$ attempts (where applicable) for the $[[506,240]]$
quasi-cyclic code on the depolarizing channel.}
\end{figure}

The FER performance and average required iterations for decoders with
$N=100$ maximum attempts are shown in Figs. \ref{n506 QC FER fig}
and \ref{n506 QC iters fig} respectively. On the previous two codes
the augmented $\mathrm{GF}(2)$ decoder gave a similar or lower FER
than the adjusted decoder. This is not the case here with the adjusted
decoder giving a significantly lower FER. This suggests that the augmented
decoder has some effect in alleviating the effect of $4$-cycles in
the code's factor graph (none of which are present when using a $\mathrm{GF}(2)$
decoder for this code). The random perturbation, EFB, and augmented
$\mathrm{GF}(4)$ decoders all perform similarly on this code. The
combined decoder performs worse than the modified $\mathrm{GF}(4)$
decoders. 

\begin{figure}
\includegraphics[scale=0.55]{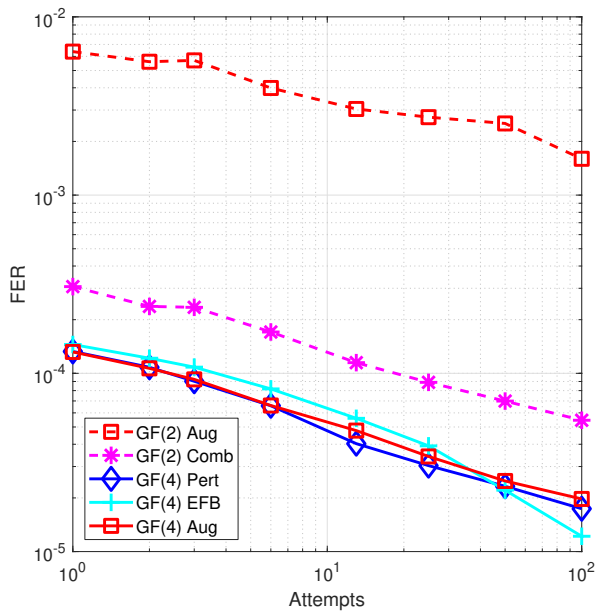}\caption{\label{n506 QC attempts fig}FER performance of decoders at $p=0.015$
with a varying number of decoding attempts for the $[[506,240]]$
quasi-cyclic code on the depolarizing channel.}
\end{figure}

Like the bicycle code, all decoding errors observed for this code
were detected errors. This is reflected in Fig. \ref{n506 QC attempts fig},
which shows an approximately linear reduction in FER with an increasing
number of maximum attempts on a log-log plot for all decoders considered. 

\subsection{Bicycle-like}

The fourth code we have considered is a $[[400,200]]$ non-dual-containing
CSS based on the bicycle-like construction of Ref. \citep{babar2016construction}.
The codes of Ref. \citep{babar2016construction} are constructed using
a BIBD in a similar way to the code of Sec. \ref{subsec:BIBD}. $\tilde{H}_{X}$
is constructed by taking the first $a$ (where $a$ is even) submatrices
of the BIBD's adjacency matrix as given in Eq. (\ref{eq:bibd adj});
that is,
\begin{equation}
\tilde{H}_{X}=\left(\begin{array}{cccc}
A_{1} & A_{2} & \cdots & A_{a}\end{array}\right).
\end{equation}
$\tilde{H}_{Z}$ is then a cyclically shifted version of $\tilde{H}_{X}$
with
\begin{equation}
\tilde{H}_{Z}=\left(\begin{array}{cccccccc}
A_{\frac{a}{2}+1} & A_{\frac{a}{2}+2} & \cdots & A_{a} & A_{1} & A_{2} & \cdots & A_{\frac{a}{2}}\end{array}\right).
\end{equation}
The use of a BIBD with $\lambda=1$ ensures that $\tilde{H}_{X}$
and $\tilde{H}_{Z}$ are both free of $4$-cycles. However, we have
observed that codes constructed in this way have low distances and
are therefore not appropriate for comparing decoders. We have found
that this distance can be increased by generalizing the construction
to allow the circulant matrices $A_{1},\dots,A_{a}$ to be randomly
generated. Note that this comes at the expense of introducing $4$-cycles.
For our code we have constructed $\tilde{H}_{X}$ from four $100\times100$
circulant matrices of weight five. Each of $\tilde{H}_{X}$ and $\tilde{H}_{Z}$
yield factor graphs with $1,700$ $4$-cycles, compared to the $2,737$
$4$-cycles of the bicycle code considered in Sec. \ref{subsec:Bicycle}. 

\begin{figure}
\includegraphics[scale=0.55]{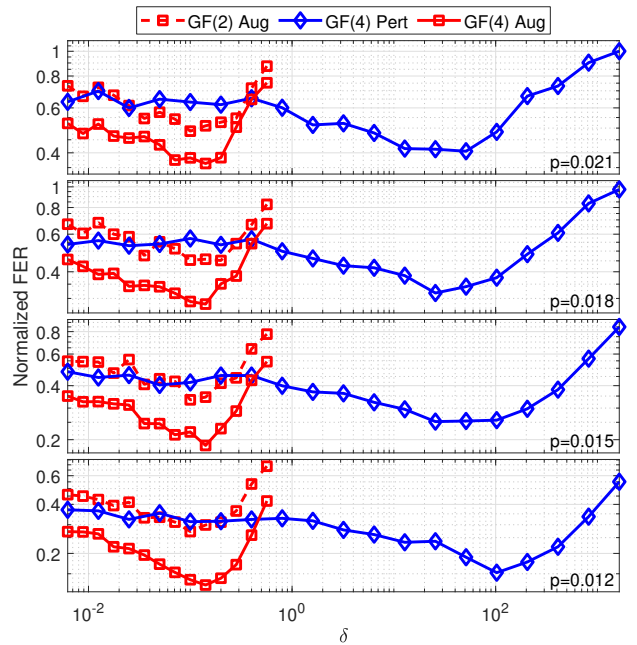}\caption{\label{n400 AR dens fig}The effect of augmentation density and random
perturbation strength on decoder performance for the $[[400,200]]$
bicycle-like code on the depolarizing channel. Each decoder uses $N=10$
maximum attempts.}
\end{figure}

The effect of augmentation density and random perturbation strength
for decoders with $N=10$ on this code is shown in Fig. \ref{n400 AR dens fig}.
Based on these results, we have selected values of $\delta=0.1$ for
the augmented $\mathrm{GF}(2)$ decoder, $\delta=0.15$ for the augmented
$\mathrm{GF}(4)$ decoder, and $\delta=100$ for the random perturbation
$\mathrm{GF}(4)$ decoder.

\begin{figure}
\includegraphics[scale=0.55]{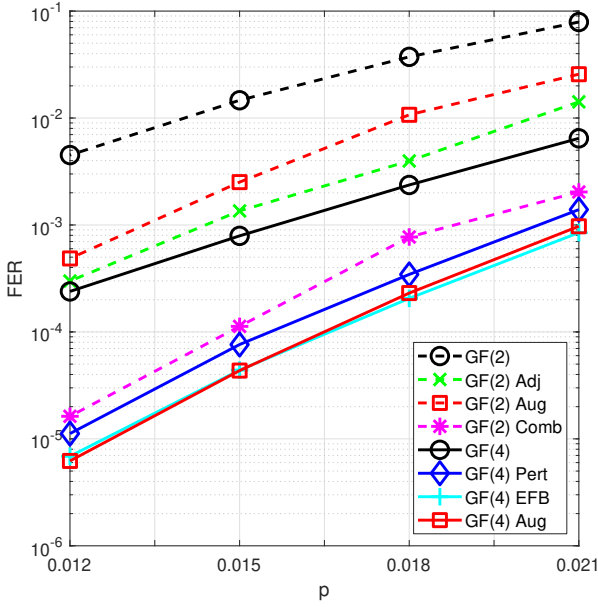}\caption{\label{n400 AR FER fig}FER performance of decoders with $N=100$
attempts (where applicable) for the $[[400,200]]$ bicycle-like code
on the depolarizing channel.}
\end{figure}

\begin{figure}
\includegraphics[scale=0.55]{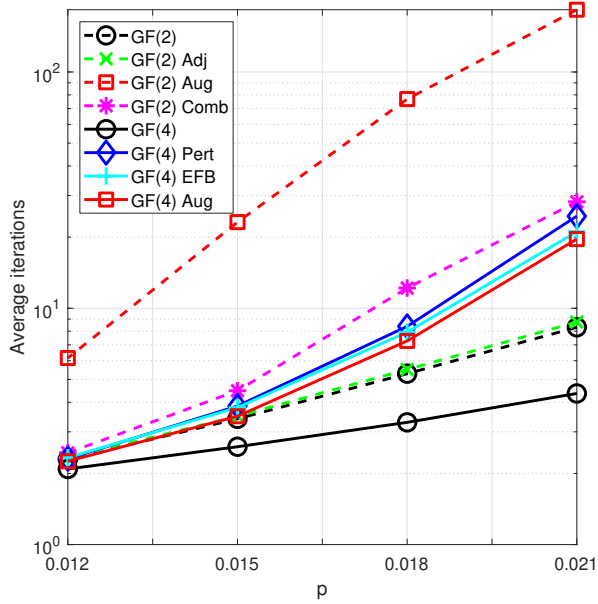}\caption{\label{n400 AR iters fig}Average number of iterations required by
decoders with $N=100$ attempts (where applicable) for the $[[400,200]]$
bicycle-like code on the depolarizing channel.}
\end{figure}

The FER performance and average required iterations for decoders with
$N=100$ maximum attempts are shown in Figs. \ref{n400 AR FER fig}
and \ref{n400 AR iters fig} respectively. Again, the adjusted decoder
outperforms the augmented $\mathrm{GF}(2)$ decoder; however, the
gap in their performance is smaller than for the quasi-cyclic code
of Sec. \ref{subsec:Quasi-cyclic}. The EFB and augmented $\mathrm{GF}(4)$
decoders perform similarly on this code, both outperforming the random
perturbation decoder. The combined decoder is again outperformed by
all modified $\mathrm{GF}(4)$ decoders, although the performance
gap is smaller than in the quasi-cyclic case.

\begin{figure}
\includegraphics[scale=0.55]{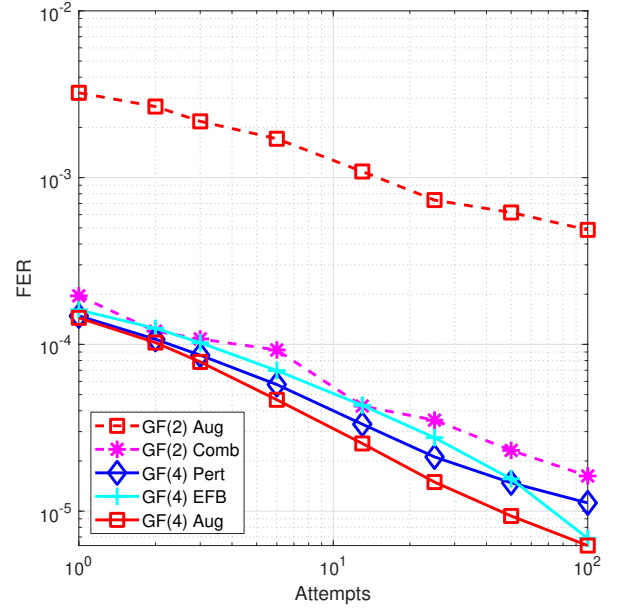}\caption{\label{n400 AR attempts fig}FER performance of decoders at $p=0.012$
with a varying number of decoding attempts for the $[[400,200]]$
bicycle-like code on the depolarizing channel.}
\end{figure}

While our modified construction gives a higher distance than the codes
presented in Ref. \citep{babar2016construction}, we still observed
a moderate number of undetected errors, which can be attributed to
the codes moderatly low distance of $d\leq10$. For example, at $p=0.015$
approximately $15\%$ of errors are undetected for both the EFB and
augmented $\mathrm{GF}(4)$ decoders. However, this is not significant
enough fraction of errors to prevent the FER reducing near-linearly
on a log-log plot with an increasing maximum number of attempts as
shown in Fig. \ref{n400 AR attempts fig}.

\subsection{Non-CSS A\label{subsec:Non-CSS-A}}

The fifth code we have considered is a $[[400,202]]$ non-CSS code
based on construction three of Ref. \citep{tan2010efficient}. The
$\mathrm{GF}(2)$ and $\mathrm{GF}(4)$ parity-check matrices for
this code are defined by the matrices $H_{X}$ and $H_{Z}$ as outlined
in Eqs. (\ref{eq:gf2 pcm}) and (\ref{eq:gf4 pcm}). For this code
these matrices are of the form
\begin{equation}
H_{X}=\left(\begin{array}{cccc}
A_{X}^{(1)} & A_{X}^{(2)} & \cdots & A_{X}^{(a)}\end{array}\right),\label{eq:ncssa Hx}
\end{equation}
\begin{equation}
H_{Z}=\left(\begin{array}{cccc}
A_{Z}^{(1)} & A_{Z}^{(2)} & \cdots & A_{Z}^{(a)}\end{array}\right).\label{eq:ncssb Hz}
\end{equation}
The submatrices $A_{X}^{(i)}$ and $A_{Z}^{(i)}$ are given by
\begin{equation}
A_{X}^{(i)}=\left(\begin{array}{cc}
B_{X}^{(i)} & B_{X}^{(i)T}P_{i}^{T}\\
P_{i}B_{X}^{(i)T} & P_{i}B_{X}^{(i)}P_{i}^{T}
\end{array}\right),
\end{equation}
\begin{equation}
A_{Z}^{(i)}=\left(\begin{array}{cc}
B_{Z}^{(i)} & B_{Z}^{(i)T}P_{i}^{T}\\
P_{i}B_{Z}^{(i)T} & P_{i}B_{Z}^{(i)}P_{i}^{T}
\end{array}\right).
\end{equation}
Here $B_{X}^{(i)}$ and $B_{Z}^{(i)}$ are square matrices of the
same size that are either both symmetric or both circulant; $P_{i}$
is a square matrix satisfying $P_{i}^{T}=P_{i}^{-1}$. For our code,
we have taken $a=2$, each $B_{X}^{(i)}$ and $B_{Z}^{(i)}$ to be
a $100\times100$ circulant matrix of weight three, and each $P_{i}$
to be a $100\times100$ permutation matrix.

\begin{figure}
\includegraphics[scale=0.55]{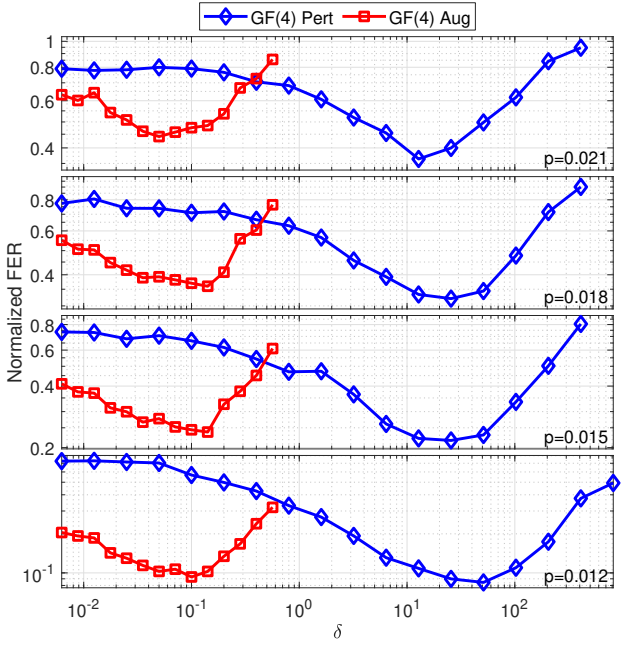}\caption{\label{n400 NCSS dens fig}The effect of augmentation density and
random perturbation strength on decoder performance for the $[[400,202]]$
non-CSS code A on the depolarizing channel. Each decoder uses $N=10$
maximum attempts.}
\end{figure}

The effect of augmentation density and random perturbation strength
for decoders with $N=10$ on this code is shown in Fig. \ref{n400 NCSS dens fig}.
Note that for non-CSS codes we can only use $\mathrm{GF}(4)$ based
decoders. Based on these results, we have chosen $\delta=0.1$ for
the augmented decoder and $\delta=25$ for the random perturbation
decoder.

\begin{figure}
\includegraphics[scale=0.55]{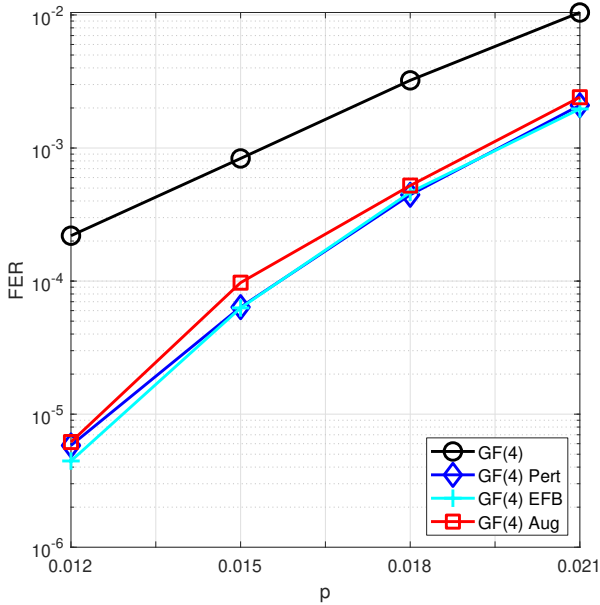}\caption{\label{n400 NCSS FER fig}FER performance of decoders with $N=100$
attempts (where applicable) for the $[[400,202]]$ non-CSS code A
on the depolarizing channel.}
\end{figure}

\begin{figure}
\includegraphics[scale=0.55]{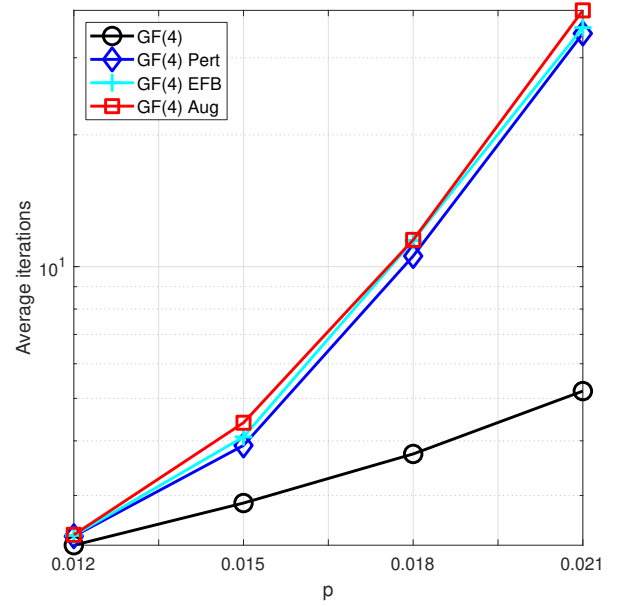}\caption{\label{n400 NCSS iters fig}Average number of iterations required
by decoders with $N=100$ attempts (where applicable) for the $[[400,202]]$
non-CSS code A on the depolarizing channel.}
\end{figure}

The FER performance and average required iterations for decoders with
$N=100$ maximum attempts are shown in Figs. \ref{n400 NCSS FER fig}
and \ref{n400 NCSS iters fig} respectively. It can be seen that the
random perturbation, EFB, and augmented decoders all perform similarly
on this code.

\begin{figure}
\includegraphics[scale=0.55]{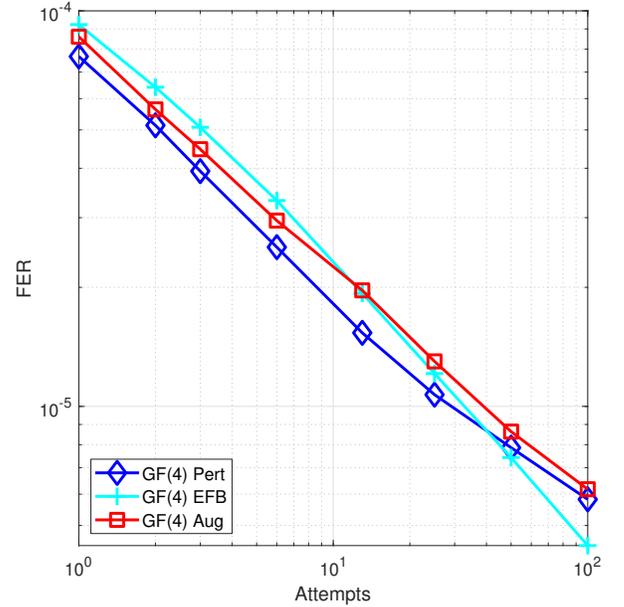}\caption{\label{n400 NCSS attempts fig}FER performance of decoders at $p=0.012$
with a varying number of decoding attempts for the $[[400,202]]$
non-CSS code A on the depolarizing channel.}
\end{figure}

The majority of decoding errors are detected errors for this code,
which also has distance $d\leq10$. At $p=0.012$ only $1-2\%$ of
errors are undetected for the random perturbation, EFB, and augmented
decoders. This is again reflected in the near-linear reduction in
FER with increasing maximum number of attempts on the log-log plot
given in Fig. \ref{n400 NCSS attempts fig}.

\subsection{Non-CSS B}

The final code we have considered is a $[[400,201]]$ non-CSS code
based on construction four of Ref. \citep{tan2010efficient}. This
construction is quite similar to that of the last section with $H_{X}$
and $H_{Z}$ defined as in Eqs. (\ref{eq:ncssa Hx}) and (\ref{eq:ncssb Hz})
respectively. However, the submatrices $A_{X}^{(i)}$ and $A_{Z}^{(i)}$
are now given by

\begin{equation}
A_{X}^{(i)}=\left(\begin{array}{cc}
B_{X}^{(i)} & B_{X}^{(i)T}P_{i}^{T}\end{array}\right),
\end{equation}
\begin{equation}
A_{Z}^{(i)}=\left(\begin{array}{cc}
B_{Z}^{(i)} & B_{Z}^{(i)T}P_{i}^{T}\end{array}\right).
\end{equation}
Here $B_{X}^{(i)}$ and $B_{Z}^{(i)}$ are either both symmetric,
both circulant, or $B_{X}^{(i)}B_{Z}^{(i)T}+B_{X}^{(i)T}B_{Z}^{(i)}$
is symmetric; $P_{i}$ is a permutation matrix. For our code we have
taken $a=1$, $B_{X}^{(1)}$ and $B_{Z}^{(1)}$ to be $200\times200$
circulant matrices of weight six, and $P_{1}$ to be a $200\times200$
permutation matrix.

\begin{figure}
\includegraphics[scale=0.55]{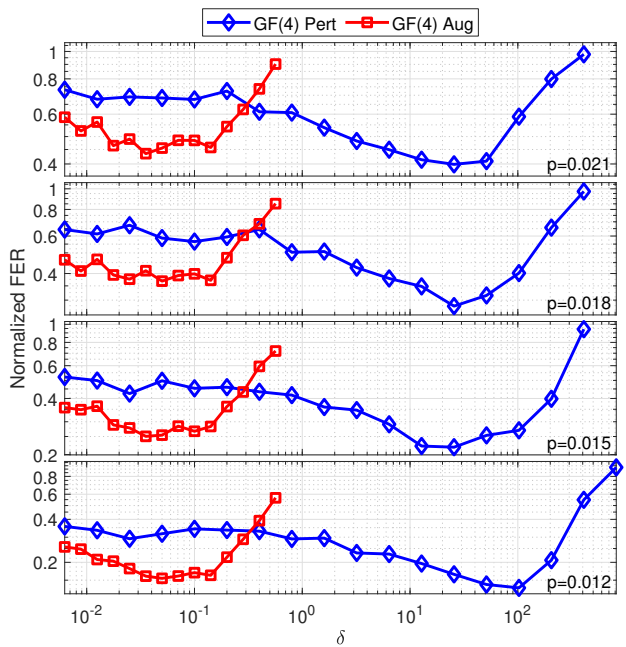}\caption{\label{n400 NCSSB dens fig}The effect of augmentation density and
random perturbation strength on decoder performance for the $[[400,201]]$
non-CSS code B on the depolarizing channel. Each decoder uses $N=10$
maximum attempts.}
\end{figure}

The effect of augmentation density and random perturbation strength
for decoders with $N=10$ on this code is shown in Fig. \ref{n400 NCSSB dens fig}.
Based on these results, we have chosen $\delta=0.05$ for the augmented
decoder and $\delta=25$ for the random perturbation decoder.

\begin{figure}
\includegraphics[scale=0.55]{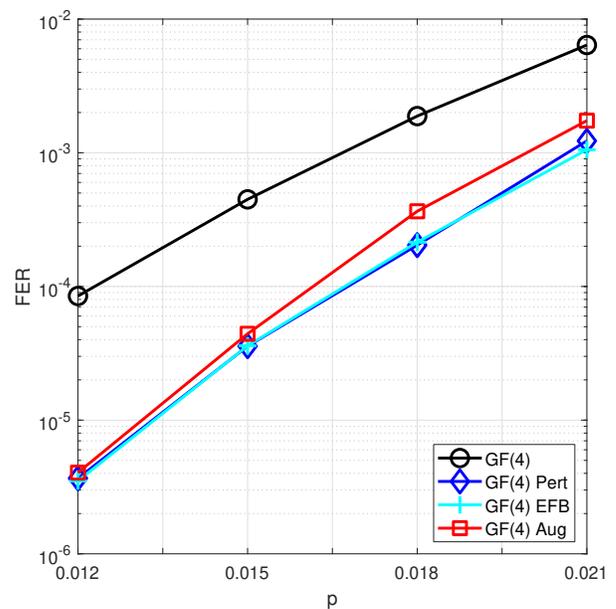}\caption{\label{n400 NCSSB FER fig}FER performance of decoders with $N=100$
attempts (where applicable) for the $[[400,201]]$ non-CSS code B
on the depolarizing channel.}
\end{figure}

\begin{figure}
\includegraphics[scale=0.55]{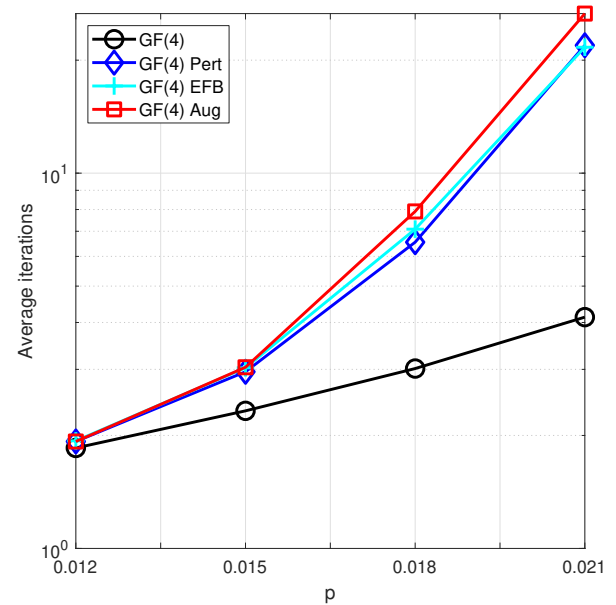}\caption{\label{n400 NCSSB iters fig}Average number of iterations required
by decoders with $N=100$ attempts (where applicable) for the $[[400,201]]$
non-CSS code B on the depolarizing channel.}
\end{figure}

The FER performance and average required iterations for decoders with
$N=100$ maximum attempts are shown in Figs. \ref{n400 NCSSB FER fig}
and \ref{n400 NCSSB iters fig} respectively. The results are consistent
with those of Sec. \ref{subsec:Non-CSS-A}, with the random perturbation,
EFB, and augmented decoders all performing fairly similarly on this
code.

\begin{figure}
\includegraphics[scale=0.55]{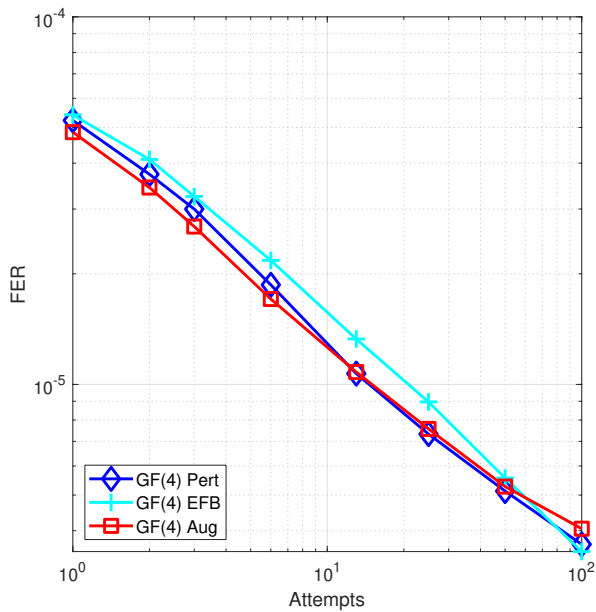}\caption{\label{n400 NCSSB attempts fig}FER performance of decoders at $p=0.012$
with a varying number of decoding attempts for the $[[400,201]]$
non-CSS code B on the depolarizing channel.}
\end{figure}

A moderate number of undetected errors were observed for this code,
which also has distance $d\leq10$. For example, at $p=0.012$ approximately
$10-15\%$ of errors are undetected for each decoder. It can also
be seen that the reduction in FER with an increasing number of maximum
attempts, while still near-linear on the log-log plot of Fig. \ref{n400 NCSSB attempts fig},
tapers of slightly more than was observed for the code of the previous
section.

\section{Conclusion\label{sec:Conclusion}}

We have presented modified belief propagation decoders for QLDPC codes
that, depending on the code, either outperform or perform similarly
to other decoders presented in literature. We have proposed the $\mathrm{GF(2)}$
based adjusted decoder, which uses modified error probabilities to
reintroduce correlations between the $X$ and $Z$ components of an
error that are lost when using a standard $\mathrm{GF}(2)$ decoder.
Furthermore, e have demonstrated that the augmented decoder, which
has previously been proposed for classical binary LDPC codes, can
be applied in the quantum case and that it can be based on an underlying
$\mathrm{GF}(2)$, $\mathrm{GF}(4)$, or supernode decoder. We have
also proposed a combination of the augmented $\mathrm{GF}(2)$ and
adjusted decoders. For the bicycle and BIBD based dual-containing
CSS codes tested, the augmented $\mathrm{GF}(4)$, augmented supernode,
and combined decoders were shown to outperform random perturbation
and EFB decoders. For the two non-dual-containing CSS codes and the
two non-CSS codes considered, augmented $\mathrm{GF}(4)$ and supernode
decoders were shown to perform similarly to random perturbation and
EFB decoders.

\bibliographystyle{apsrev4-1}
\bibliography{sparsePaper}

\appendix

\section{Check node Fourier transform implementations}

\subsection{Classical decoding\label{sec:Belief-propagation-Fourier}}

The check constraint of Eq. (\ref{eq:check to error message}) can
be written as
\begin{equation}
\sum_{j'\in\mathcal{M}(i)\backslash j}H_{ij'}e_{j'}=\sum_{j'\in\mathcal{M}(i)\backslash j}\tilde{e}_{j'}=z_{i}-H_{ij}a,
\end{equation}
where $\tilde{e}_{j'}=H_{ij'}e_{j'}$. $\tilde{e}_{j'}$ can be used
to define
\begin{equation}
\tilde{\lambda}_{i\rightarrow j}^{a}=\sum_{\boldsymbol{e}:\sum_{j'}\tilde{e}_{j'}=a}\prod_{j'}\mu_{j'\rightarrow i}^{e_{j'}}=\sum_{\boldsymbol{e}:\sum_{j'}\tilde{e}_{j'}=a}\prod_{j'}\tilde{\mu}_{j'\rightarrow i}^{\tilde{e}_{j'}},\label{eq:check node convolution}
\end{equation}
where $\tilde{\mu}_{j'\rightarrow i}^{\tilde{e}_{j'}}=\mu_{j'\rightarrow i}^{H_{ij'}^{-1}\tilde{e}_{j'}}$
(this corresponds to a permutation of elements) and $j'\in\mathcal{M}(i)\backslash j$.
Eq. (\ref{eq:check node convolution}) is a convolution and as such
can be efficiently computed using a Fourier transform $\mathcal{F}$
as
\begin{equation}
\tilde{\lambda}_{i\rightarrow j}=K\mathcal{F}^{-1}\left\{ \prod_{j'}\mathcal{F}\{\tilde{\mu}_{j'\rightarrow i}\}\right\} ,
\end{equation}
where $\mathcal{F}^{-1}$ is the inverse Fourier transform and the
product is element-wise ($K$ is a normalization factor). A Hadamard
transform can be used in the binary case; if $\tilde{\mu}_{j'\rightarrow i}=(\tilde{\mu}_{j'\rightarrow i}^{0},\tilde{\mu}_{j'\rightarrow i}^{1})$
is considered as a column vector, then
\begin{equation}
\mathcal{F}\{\tilde{\mu}_{j'\rightarrow i}\}=F\tilde{\mu}_{j'\rightarrow i},
\end{equation}
where
\begin{equation}
F\propto\left(\begin{array}{cc}
1 & 1\\
1 & -1
\end{array}\right).
\end{equation}
The inverse transform is also achieved through multiplication by $F$
(up to some unimportant scaling factor). $\lambda_{i\rightarrow j}$
is a permuted version of $\tilde{\lambda}_{i\rightarrow j}$ with
\begin{equation}
\lambda_{i\rightarrow j}^{a}=\tilde{\lambda}_{i\rightarrow j}^{z_{i}-H_{ij}a}.
\end{equation}

\subsection{$\mathrm{GF}(4)$ stabilizer decoding\label{sec:Stabiliser-code-Fourier}}

The check constraint of Eq. (\ref{eq:quantum check node}) can be
written as
\begin{equation}
\mathrm{tr}(H_{ij}\bar{a}+\sum_{j'\in\mathcal{M}(i)\backslash j}H_{ij'}\bar{e_{j'}})=\mathrm{tr}(H_{ij}\bar{a}+\sum_{j'\in\mathcal{M}(i)\backslash j}\tilde{e}_{j'})=z_{i},
\end{equation}
where $\tilde{e}_{j'}=H_{ij'}\bar{e_{j'}}$. $\tilde{\lambda}_{i\rightarrow j}^{a}$
is defined in the same was as Eq. (\ref{eq:check node convolution})
with $\tilde{\mu}_{j'\rightarrow i}^{\tilde{e}_{j'}}=\mu_{j'\rightarrow i}^{(H_{ij'}^{-1}\tilde{e}_{j'})^{-1}}=\mu_{j'\rightarrow i}^{H_{ij'}\tilde{e}_{j'}^{-1}}$.
Again, $\tilde{\lambda}_{i\rightarrow j}$ can be calculated using
the Hadamard transform with
\begin{equation}
F\propto\left(\begin{array}{cc}
1 & 1\\
1 & -1
\end{array}\right)^{\otimes2}=\left(\begin{array}{cccc}
1 & 1 & 1 & 1\\
1 & -1 & 1 & -1\\
1 & 1 & -1 & -1\\
1 & -1 & -1 & 1
\end{array}\right).
\end{equation}
$\tilde{\lambda}_{i\rightarrow j}^{a}$ corresponds to $\sum_{j'}\tilde{e}_{j'}=a$
and as such can be used to determine $\lambda_{i\rightarrow j}^{a}$,
which corresponds to $e_{j}=a$. If $z_{i}=0$, then $H_{ij}\bar{a}+\sum_{j'}\tilde{e}_{j'}=0$
or $1$; conversely, if $z_{i}=1$, then $H_{ij}\bar{a}+\sum_{j'}\tilde{e}_{j'}=\omega$
or $\bar{\omega}$. Therefore, for $z_{i}=0$ the elements of $\lambda_{i\rightarrow j}$
are
\begin{equation}
\lambda_{i\rightarrow j}^{a}=\frac{1}{2}\left[\tilde{\lambda}_{i\rightarrow j}^{-H_{ij}\bar{a}}+\tilde{\lambda}_{i\rightarrow j}^{1-H_{ij}\bar{a}}\right],
\end{equation}
and for $z_{i}=1$
\begin{equation}
\lambda_{i\rightarrow j}^{a}=\frac{1}{2}\left[\tilde{\lambda}_{i\rightarrow j}^{\omega-H_{ij}\bar{a}}+\tilde{\lambda}_{i\rightarrow j}^{\bar{\omega}-H_{ij}\bar{a}}\right].
\end{equation}
These can be combined to give
\begin{equation}
\lambda_{i\rightarrow j}^{a}=\frac{1}{2}\left[\tilde{\lambda}_{i\rightarrow j}^{\omega z_{i}-H_{ij}\bar{a}}+\tilde{\lambda}_{i\rightarrow j}^{\omega z_{i}+1-H_{ij}\bar{a}}\right].
\end{equation}

\end{document}